\pdfoutput=1
\documentclass[12pt,preprint]{aastex}

\newcommand{\gps}{\ensuremath{g_{\rm P1}}}
\newcommand{\rps}{\ensuremath{r_{\rm P1}}}
\newcommand{\ips}{\ensuremath{i_{\rm P1}}}
\newcommand{\zps}{\ensuremath{z_{\rm P1}}}
\newcommand{\yps}{\ensuremath{y_{\rm P1}}}
\newcommand{\wps}{\ensuremath{w_{\rm P1}}}

\newcommand{\PS}{\protect \hbox {Pan-STARRS1}}

\usepackage{verbatim}
\usepackage{amsmath}

\def\deg{\ifmmode^\circ\else$^\circ$\fi}
\def\arcsec{\ifmmode^{\prime\prime}\else$^{\prime\prime}$\fi}
\def\arcmin{\ifmmode^{\prime}\else$^{\prime}$\fi}
\def\figwid{4.0in}

\shorttitle{Pan-STARRS Photometric System}
\shortauthors{J.L. Tonry et al.}
\begin{document}
\title{The Pan-STARRS1 Photometric System}
%
%
%
\author{J.L.~Tonry,\altaffilmark{1}
C.W.~Stubbs,\altaffilmark{2,3}
K.R.~Lykke,\altaffilmark{4}
P.~Doherty,\altaffilmark{3}
I.S.~Shivvers, \altaffilmark{2,5}
W.S.~Burgett,\altaffilmark{1}
K.C.~Chambers,\altaffilmark{1} 
K.W.~Hodapp,\altaffilmark{1}
N.~Kaiser,\altaffilmark{1}
R.-P.~Kudritzki,\altaffilmark{1}
E.A.~Magnier,\altaffilmark{1}
J.S.~Morgan,\altaffilmark{1}
P.A.~Price,\altaffilmark{7} and
R.J.~Wainscoat \altaffilmark{1}
}


\altaffiltext{1}{Institute for Astronomy, University of Hawaii, 2680
  Woodlawn Drive, Honolulu HI 96822}
\altaffiltext{2}{Harvard-Smithsonian Center for Astrophysics, 60
  Garden Street, Cambridge, MA 02138} \altaffiltext{3}{Department of
  Physics, Harvard University, 17 Oxford Street, Cambridge MA 02138}
\altaffiltext{4} {National Institute of Standards and Technology, 100 Bureau Drive, Gaithersburg MD 20899, USA}
\altaffiltext{5}{Department of Astronomy, University of California, Berkeley CA 94720}
\altaffiltext{6}{Department of Physics and Astronomy, Johns Hopkins
  University, 3400 North Charles Street, Baltimore, MD 21218, USA}
\altaffiltext{7}{Department of Astrophysical Sciences, Princeton
  University, Princeton, NJ 08544, USA} 
 \altaffiltext{8}{US Naval Observatory, Flagstaff Station, Flagstaff, AZ 86001, USA}


\begin{abstract}
The \PS\ survey is collecting multi-epoch, multi-color observations of
the sky north of declination $-30\deg$ to unprecedented depths.  These
data are being photometrically and astrometrically calibrated and will
serve as a reference for many other purposes.  In this paper we
present our determination of the \PS\ photometric system: \gps, \rps,
\ips, \zps, \yps, and \wps.  The \PS\ photometric system is
fundamentally based on the HST Calspec spectrophotometric
observations, which in turn are fundamentally based on models of white
dwarf atmospheres.
We define the \PS\ magnitude system, and describe in detail our
measurement of the system passbands, including both the instrumental
sensitivity and atmospheric transmission functions.  Byproducts,
including transformations to other photometric systems, galactic
extinction, and stellar locus are also provided.  We close with a
discussion of remaining systematic errors.
\end{abstract}


\vfil
\eject
\clearpage


\section{INTRODUCTION}
\label{sec:intro}

\subsection{Photometry and Astronomy}

All ground-based photometry measures light that has been filtered by passage
through the atmosphere and by an optical system that typically
includes a bandpass filter.  The surviving light is finally converted
into an electrical signal by a detector.  The system therefore
presents a net capture cross section, $A(\nu,\theta,t)$, to incoming
photons that depends on frequency $\nu$ (or wavelength), direction
$\theta$ with respect to the boresight (or detector pixel), and time,
where ``capture cross section'' quantifies the probability of counting
an incident photon as an $e^-$ in the detector.

An object with a spectral energy distribution (SED)
$f_\nu$~[erg/s/cm$^2$/Hz] whose light arrives at the top of the
atmosphere therefore creates a signal of $\int f_\nu\, (h\nu)^{-1}\,
A(\nu,\theta,t)\,d\nu$ in a photon-sensitive detector.  If the
instrument's bandpass encompasses significant wavelength variation in
$A(\nu,\theta,t)$ or $f_\nu$ it is {\it not} possible to recover
$f_\nu$ uniquely from an observation: information is necessarily lost,
and different SEDs can produce the same signal.  In many cases we are
interested in a restricted question, however; we believe we know the
spectral form of the SED of an object, but we do not know the overall
normalization.  In this case we can recover this normalization by
simply integrating a unity-normalized SED against the known cross
section $A(\nu,\theta,t)$, and then scale to the true SED by the ratio
of the observed signal to this integral.

Astronomical magnitude systems are based on this concept, as
summarized by \cite{Bessel05}.  The ``Vega normalized'' system,
developed when instrumentation was capable of much higher relative
accuracy than absolute, uses A0 stars (e.g. Vega) as the reference.
That is, a ``Vega magnitude'' is the ratio of the signal produced by
integrating an object's SED through $A(\nu,\theta,t)$ compared to the
A0 star Vega, where ``A0 star'' has evolved to a loosely defined set
of stars whose SEDs are believed to be known at the few percent level,
and whose cataloged magnitudes are fairly self consistent with the
SEDs.  (In retrospect, the choice of bright A0 stars with enormous H
absorption for the standard SED was less than optimal.)  This
magnitude system is therefore operationally defined from a catalog as
opposed to physically defined, and systematic inaccuracies accrue from
the definition as well as from uncertain knowledge of bandpasses and
detector sensitivities.

Another magnitude system, heartily endorsed by \PS, is the ``AB
system'' \citep{Oke+Gunn83}, described in detail for the Sloan Digital
Sky Survey (SDSS, \cite{SDSS}) by \cite{Fukugita96}.  In this system a
``monochromatic AB magnitude'' is just a logarithm of flux density:
\begin{align}
m_{AB}(\nu) &= -2.5\log(f_\nu/3631~\hbox{Jy})\\
 &= -48.600 - 2.5\log(f_\nu[\hbox{erg/sec/cm$^2$/Hz])}\\
 &= 16.847 - 2.5\log(f_\gamma[\hbox{ph/sec/cm$^2$/d$\ln\lambda$])}.
\end{align}
where $1~\hbox{Jy} = 10^{-23}\hbox{erg/sec/cm$^2$/Hz}$, $f_\gamma$
only differs from $f_\nu$ by a factor of $h$ but is integrated against
d$\ln\lambda = d\ln\nu$ without needing a factor of $(h\nu)^{-1}$, and
the constant was chosen to set the AB mag of Vega at 548~nm to be
0.03, the $V$ mag of Vega, under the assumptions that the ``effective
wavelength'' of the V band for Vega was 548~nm.  A ``bandpass AB
magnitude'' is defined similarly:
\begin{equation}
m_{AB} = -2.5\log\left(\int f_\nu\,(h\nu)^{-1}\,A(\nu)\, d\nu \over
\int 3631~\hbox{Jy}\,(h\nu)^{-1}\,A(\nu)\, d\nu \right).
\end{equation}
There is no arbitrariness in the magnitude definition for a
given well-defined capture cross section $A(\nu)$\footnote{The classic
  observer's ``magnitude'' system, originally defined by Pogson to
  crudely coincide with ancient Greek classification of star
  brightness, is slowly withering in favor of flux densities reported
  in units of Jy, but we caution that such flux densities typically
  {\it are} ambiguous for extended bandpasses, and we {\it strongly}
  recommend that non-monochromatic ``flux densities'' conform to this
  definition of the AB system: A non-monochromatic ``flux density'' is
  the ratio of detector response to SED relative to constant $f_\nu$.}.

%
%
%
%

In practice we do not burden every flux observation with a detailed
$A(\nu,\theta,t)$, so we instead alter the flux we report to reflect
the flux we think it would have had if subjected to a nominal $A(\nu)$
instead of the actual, momentary $A(\nu,\theta,t)$.  This correction
has a number of components, including removing the dependence on
$\theta$ by correcting for detector response, optics vignetting and
other spatial variations, and most significantly, adjusting the flux
for the instantaneous ``atmospheric extinction''.  The
$\nu$-independent component of this amounts merely to a
renormalization of sensitivity; the $\nu$-dependent component creates
a new source of error that depends on the particular SED being
observed, generally quantified as a ``color term'', a
correction for stellar SEDs consisting of a coefficient that
multiplies a color.

Because of the ambiguity when inferring an AB magnitude from a flux
measurement when the SED is not known, there is a choice to be made
between trying to report an AB magnitude that is ``universal''
(e.g. what would be observed without any atmospheric extinction)
versus what is actually observed.  The tradition of ``regression to
top of atmosphere'' makes sense for a limited set of SEDs such as
stars, but non-stellar SEDs (e.g. very cool stars, AGN at high
redshift, supernovae, etc.) are becoming so important, that \PS\ has
adopted the approach of modern photometric systems such as 
SDSS: bandpasses explicitly include a
nominal level of atmospheric extinction.\footnote{\PS\ does {\it not},
  however, adhere to the SDSS practice of reporting inverse hyperbolic
  sines (luptitudes) in place of magnitudes; this practice arose from
  attempting to serve two priors (object power, for which a logarithm
  is appropriate, versus net observed flux, for which linear is
  better) with one number.  The \PS\ databases simply serve up both
  flux and magnitude.}

The actual determination and implementation of the $AB$ magnitude
system for \PS\ can be carried out in different ways.  One option is
to infer AB magnitudes through synthetically derived bandpasses by
various methods such as 1) exploiting the overlap between \PS\ and
another extensive catalog of stellar magnitudes such as SDSS, 2) using
the stellar locus in color-color space (subject to removal of dust
reddening), or 3) observing spectrophotometric standard stars and
regressing out atmospheric extinction.  In effect this transfers an
extant $AB$ calibration (for better or for worse) to \PS\ photometry.

Another option (described in \cite{PrecisionPhot}) is to obtain
independent determinations of the instrumental and atmospheric
response functions, establishing the wavelength-dependent part of
$A(\nu)$.  In principle it is possible to determine the absolute
sensitivity of the \PS\ system, but in practice we can also depend on
spectrophotometric standard star observations to verify the bandpasses
and set the overall normalization.

Once a set of stars are provided with accurate AB magnitudes, the
system can be propagated around the sky using overlapping observations
to disentangle instrumental and atmospheric contributions.  This was
used successfully by \cite{Padmanabhan08} for the SDSS survey and is
currently being implemented for \PS\ by \cite{Schlafly12}.

The consistency of these different techniques can be used to assess
systematic errors in the survey's photometric calibration, but we
defer a detailed comparison of these different approaches to a
subsequent paper. Our initial comparisons indicate that the
\PS\ implementation of the $AB$ system has an accuracy of $\sim0.02$
mag (90\% confidence), where the dominant contribution is uncertainty
in how well spectrophotometry matches the AB system.

For this paper we use a combination of measurements of our
instrumental and atmospheric response function with spectrophotometric
star observations to establish the \PS\ photometric system. The
\PS\ calibration described here is fundamentally based upon the
Calspec spectrophotometric standards from Hubble Space Telescope (HST)
\citep{Calspec}.

\subsection{\PS}

The \PS\ system is a 1.8~m aperture, f/4.4 telescope
\citep{PS1_optics} illuminating a 1.4~Gpixel detector spanning a
3.3\deg\ field of view (\cite{GPC} and \cite{StarGrasp}), located on
Haleakala (\cite{PS1_system}), and dedicated to sky survey
observations (\cite{PS_MDRM}).  The \PS\ filters are designated
\gps, \rps, \ips, \zps, \yps, and \wps\ in order to clearly
distinguish PS1 from other photometric systems.  The gigapixel camera
(GPC1) consists of an $8\times8$ array of orthogonal transfer array
(OTA) CCDs, and each OTA is subdivided into an $8\times8$ array of
``cells'', each an independent $590\times598$ 10$\mu$m pixel CCD.
Images obtained by the \PS\ system are processed through the Image
Processing Pipeline (IPP) \citep{PS1_IPP}.  Although the filter
system for \PS\ has much in common with that used in previous surveys
such as SDSS \citep{SDSS}, the \gps\ filter extends 20~nm redward of
$g_{SDSS}$, paying the price of 5577\AA\ sky emission for greater
sensitivity and lower systematics for photometric redshifts, the
\zps\ filter is cut off at 920~nm, giving it a different response than
the detector response defined $z_{SDSS}$, and SDSS has no
corresponding \yps\ filter.

AB magnitudes reported by \PS\ include an explicit model for the
atmospheric extinction at a nominal airmass (1.2), with relatively
small corrections applied to the observed fluxes to bring them to this
airmass.  Given a known SED that differs from $f_\nu$=const, it is
therefore possible to convert the ``top of atmosphere'' magnitude
reported by \PS\ back to the nominal airmass, and from that correct
for the atmospheric extinction for the particular SED.
This correction may be small or parameterizable by ``color terms'' if
the SED is similar to that of a star, but it can be very significant
for an object with an emission line in an atmospheric absorption band.

As described at length by \cite{PrecisionPhot}, \cite{PASPatmos}, and
\cite{lasercal} we believe that it is possible, at least in principle,
to calibrate the \PS\ system as an precise photometer, permitting
measurement of absolute fluxes with no reliance on standard stars
whatsoever.  Although we currently fall distinctly short of this
ideal, the next section describes our progress in implementing such a
calibration.

The third section presents the \PS\ system: \PS\ bandpasses, derived
quantities such as conversions to other bandpasses, Galactic
extinction in the \PS\ bandpasses, the stellar locus in
\PS\ bandpasses, and color terms from filter non-uniformities and
detector differences.

The penultimate section discusses sources of systematic error such as
uncertainties in bandpasses, imperfect knowledge of the atmosphere,
and uncertainties in flux determination.  We also describe some of the
foundational systematic errors such the accuracy with which SEDs do
match the AB system and point out inconsistencies between the \PS\
and SDSS photometric systems.

We conclude with an assessment of the present state of photometric
accuracy, the \PS\ strategy for carpeting the sky with photometry
accurate to better than 1 percent, and the next steps towards our
goal of photometry based on NIST calibrated equipment rather than
standard stars.

\section{PHOTOMETRIC CALIBRATION OF \PS}
\label{sec:transmission}

We factor the \PS\ system's cross section $A(\nu,\theta,t)$ into three
terms: the ``throughput'' of the optics and detector combination
(common to observations made in all filters), the filter bandpasses,
and the atmosphere.  Each of these terms presents unique challenges
for measurement and for monitoring since they change on different
timescales.

We have measured each of these factors, and in our judgement the
best information we have comes from
\begin{itemize}
\item{} Throughput, including transmission of optics and QE of
  detector: obtained by a comparison against a calibrated photodiode
  as a function of wavelength, normalized and tweaked by 3\% RMS into
  agreement with spectrophotometric standard stars.
\item{} Transmission of filters: manufacturer's measurements verified
  by in-situ measurement, tweaked by 1\% RMS into agreement with
  spectrophotometric standard stars. 
\item{} Atmospheric transmission: MODTRAN models, aerosol extinction
  tweaked into agreement with nightly regression against airmass.
\end{itemize}

This approach exploits external information for those aspects of
$A(\nu)$ that have rapid spectral variation (filter edges and
atmospheric absorption lines), while using the standard star
observations to establish the overall normalization across the
different filter bands, by applying a low order ``tweak'' to $A(\nu)$
to achieve agreement between synthetic and observed photometry.
 
This section presents a brief synopsis of the measurement of the
instrumental and atmospheric response factors, and the details the
procedure we used for bringing them into agreement with Calspec
spectrophotometry.

\subsection{Instrumental Transmission}

We have previously described \citep{lasercal} an in-dome determination
of the \PS\ filters and instrumental sensitivity function, but we
undertook new measurements that avoid the unwanted contributions from
scattered light and from ghosting in the optical system.  We used an
Ekspla laser that can be tuned from 400~nm to 1100~nm, wavelength
calibration checked against an emission line source using a
spectrometer.  The laser light was transmitted by a fiber into a 75~mm
projection telescope that created a spotlight beam we fed into
the \PS\ optics.  A NIST-calibrated photodiode was placed directly in
the projection telescope beam in order to measure the relative photon
flux as a function of wavelength; no attempt was made to measure the
absolute flux or collecting area of \PS.

The beam from the projection telescope was placed near the mean radius
of the \PS\ pupil, angled down the boresight to avoid all scatterers
and arrive at the center of GPC1, and defocused slightly to create
spots of diameter 1.0\deg\ or 1.7\deg\ that span many OTAs.  The basic
exposure comprises opening the GPC1 shutter, opening the laser shutter
and integrating the total light received by the photodiode, closing
the laser shutter when a predetermined level has been received, and
then closing the GPC1 shutter.  Observations of this sort are
interleaved with ``dark'' exposures of identical duration but without
any laser light, and the ``dark'' levels are subtracted from both the
signal from GPC1 and the photodiode.  The resulting images were then
``flatfielded'' by dividing by the gain for each cell's amplifier,
creating images whose values are the $e^-$ created in each cell.
(GPC1 includes a deployable $^{55}$Fe source inside the cryostat; a
$K\alpha$ xray photon converts to 1620 electrons in silicon; analysis
of these events provides the conversion between $e^-$ to ADU.)

Many sets of observations were collected between 400~nm and 1100~nm in
2~nm steps through each filter and with no filter.  Normalized by the
photodiode, the ratio of filter to no filter immediately provided a
measure of filter throughput, albeit in nearly parallel light at a
fairly representative angle off of normal.  The normalized fluxes with
no filter in the beam give us the optics and detector throughput, once
they are corrected for ghosts and scattering of IR light within the
silicon, both of which remove light from a small PSF but leave it
within the projected spot.

The \PS\ optical system has a particularly important ghost that is
created by light bouncing off of the CCD, the back surface of the
first corrector lens, and then being nearly refocussed onto the
detector again.  We exploited the shadow of the photodiode in the
projected spot to evaluate the ghost amplitude as well as observations
where the projected spot was focussed to 0.3\deg\ and put off center.
We found the ghost amplitude to be somewhat larger than expected from
the manufacturer's estimates of reflectivities of CCD and AR
coating on the lens, particularly at the very blue and red ends of the
spectrum.  We speculate that this arises because of one or more
imperfect or degraded lens AR coatings, and this is the reason that the
in-situ measurement of throughput is so important.

At wavelengths approaching the bandgap Si becomes more and more
transparent, so some fraction of light passes fully through the
collecting volume, and is absorbed after one or more reflections.
This signal is accumulated over a very large (mm scale) skirt, but
does not contribute usefully to the signal to noise of a tiny PSF and
therefore should not be included as part of the system throughput.
\cite{TBAD97} discusses of how this effect was important in the $I$
band for the SBF survey; it was also the reason that SDSS elected to
use thick CCDs for the $z$ band \citep{SDSScam}.  For the 75~$\mu$m
thick \PS\ detectors it becomes important in the \yps\ filter, and we
have applied a semi-empirical correction to the \PS\ throughput curve
to account for it.

The QE of silicon depends on detector temperature at very red
wavelengths, and we also took the opportunity to measure the detector
sensitivity as a function of temperature between 140~K and 190~K, from
which we constructed temperature coefficients for the \yps\ filter.
(At 1~$\mu$m we find the relative QE changes by about +0.3\%/K at
190~K.)

For comparison with the in-situ throughput measurement we assembled
manufacturer's measurements of reflectivities of the primary and
secondary mirrors, the transmission of the three corrector lens AR
coatings, and the QE measurements performed in the lab for each of the
CCDs scaled to a common temperature of 193~K.

\begin{figure}[htbp]
\begin{center}
\centerline{\includegraphics[width=\figwid]{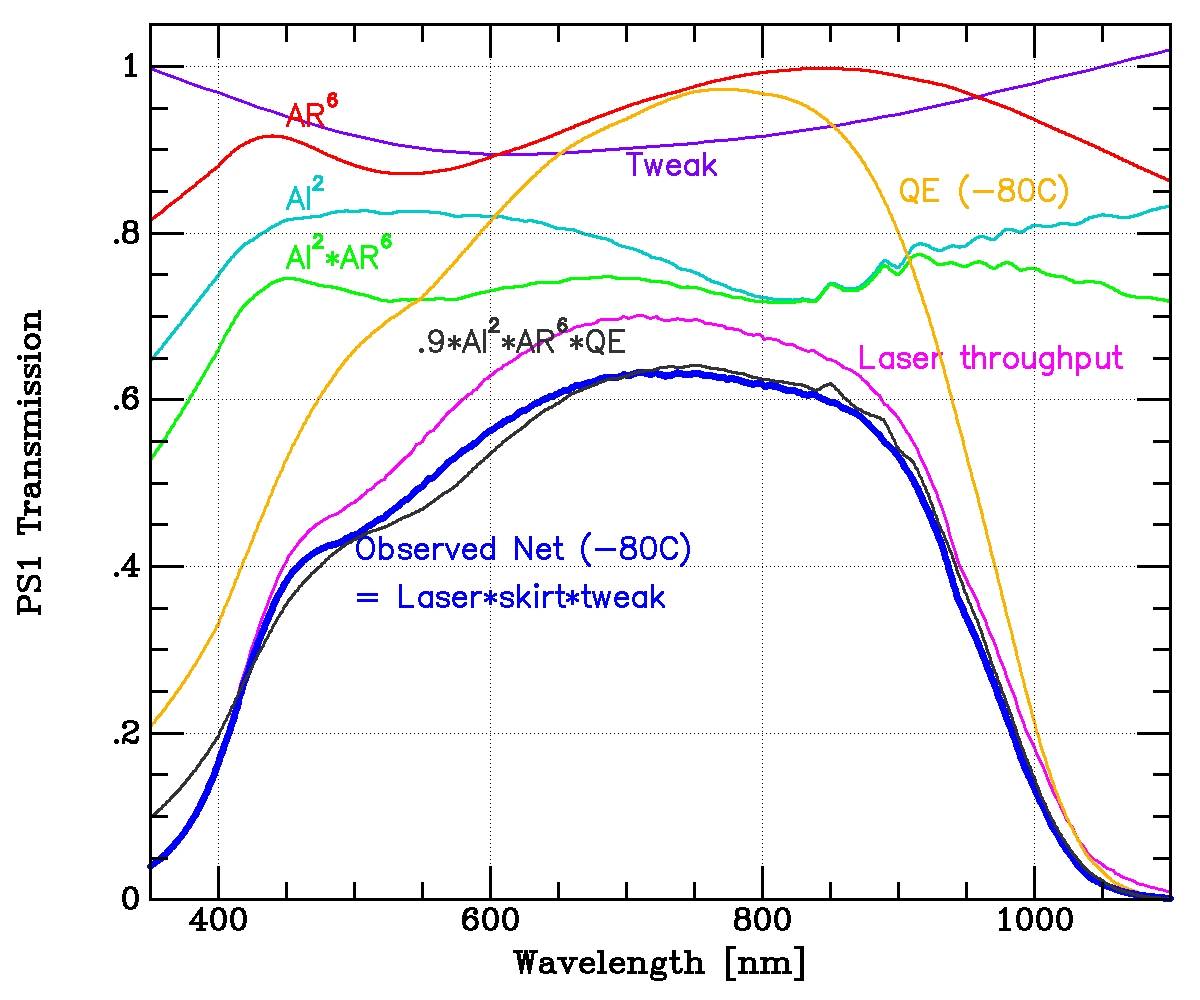}}
\caption{The various components of the relative throughput (detected
  electrons per incident photon) of the \PS\ optical system and
  detector are shown.  The heavy blue line, ``Laser throughput'' times
  the ``Tweak'' times a correction for IR skirt, is our best estimate
  of the \PS\ throughput.  The adjacent gray line,
  0.9$\times$Al$^2$AR$^6$QE, is an alternative estimate.  The
  differences we believe are the result of AR coating degradation.
  Values below 400nm are extrapolations, but no \PS\ filter has a
  significant response there.}
\label{fig:ps1trans}
\end{center}
\end{figure}

Figure~\ref{fig:ps1trans} illustrates the good agreement between the
in-situ measurements and the product of these lab throughputs.  We
have no absolute transmission normalization from our in-situ
measurements, so its absolute level is set by comparison with standard
stars.  The ``tweak'' that we use to bring the in-situ, laser
measurements into agreement with photometric standards is described
below.  Multiplying the lab throughputs by the \PS\ aperture of 1.8~m
and mean geometrical transmission past secondary and baffles of 0.62,
we found that we could match the flux detected from standard stars
if we incorporated an additional factor of $\sim0.9$, very plausibly the
result of absorption or scattering by dust and degradation.

The spot sizes of the in-situ measurements were chosen to have small
variation in vignetting and therefore require negligible flatfield
correction.  For on-sky observations, the IPP corrects for small-scale
spatial non-uniformities by dividing by images off of the
flatfield screen.  Large-scale non-uniformities are corrected using
observations of stars dithered widely across the field of view during
times of constant atmospheric extinction.  We thus reduce
$A(\nu,\theta,t)$ to $A(\nu,t)$, at least for an SED that is
approximately that of a late K star.  We detail below color terms for
other SEDs.

\subsection{Filter Transmission}

The \PS\ filters, interference coatings on 1~cm of fused silica
manufactured by Barr Precision Optics (now Materion), are located
0.4~m above the focal plane.  Barr provided transmission measurements
using an f/8 beam at 10 radii ranging from 1 to 9.5 inches and 8
azimuths at the 9.5 inch radius.  Some of the filters have substantial
variation in transmission as a function of radius, although they
appear to have a high degree of azimuthal symmetry.  Although the
pupil is a $\sim$100~mm diameter donut on the filters, color
differences arise as a function of position.

\begin{figure}[htbp]
\begin{center}
\centerline{\includegraphics[width=\figwid]{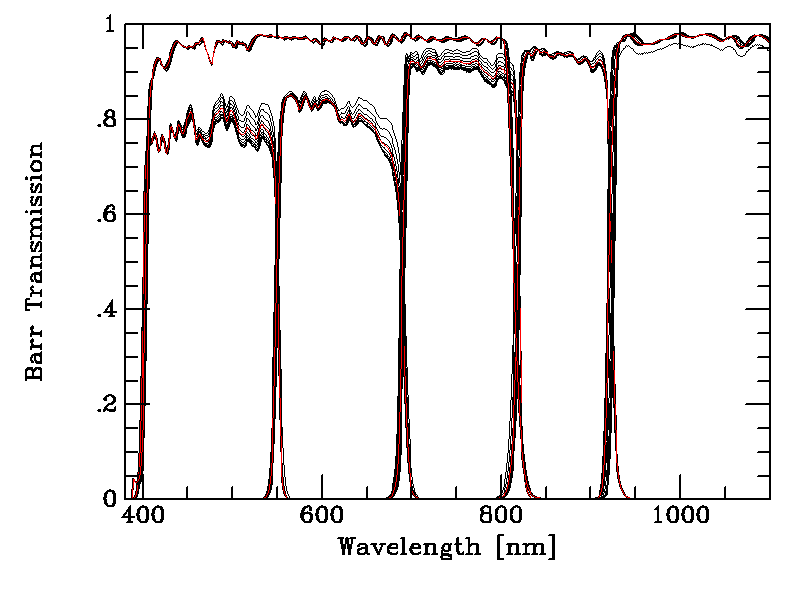}}
\caption{Filter transmission of the six \PS\ filters. \gps, \rps,
  \ips, \zps, \yps, and \wps\ are shown as a function of field angle,
  in 0.15\deg\ steps to 1.65\deg, and the red curve shows the area
  weighted average.  Small field angles tend to have similar
  transmissions, allowing their curves to be distinguished from
  large field angle.}
\label{fig:barr}
\end{center}
\end{figure}

The f/4.4 \PS\ beam is incident on the filters at angles up to
6.5\deg\ off of normal, with a pupil-averaged angle of 5.4\deg.  This
leads to a shift in transmission to the blue in the \PS\ beam relative
to Barr's nearly parallel-light data by $(1 - \sin^2 \theta /
n^2)^{1/2}$.  Calculations of Barr filter transmission at 0.0\deg\ and
9.9\deg\ off of normal provided an accurate coefficient for the
wavelength shift of 0.48\% at an angle of incidence of 9.9\deg.  For
each of the six filters and 12 field positions we ray-traced 10,000
positions across the PS1 pupil, and added up the Barr traces with
appropriate wavelength shift as a function of incident angle.  We
finally summed up a grand average that is the area-weighted
transmission out to field angles of 1.5\deg.  This is illustrated in
Figure~\ref{fig:barr}.


Our in-situ measurements of the Barr filters confirmed the accuracy of
the Barr traces of the filters.  The overall transmission and filter
edges as well as the spectral bumps and wiggles are matched to a
very satisfactory degree.  As with the throughput measurement,
however, we describe below percent-level tweaks that we require
to match standard star observations.

\subsection{Atmospheric Transmission}

The third component of the \PS\ photometric system is the atmosphere.
As described in \cite{PASPatmos}, \cite{CTIOatmos}, and
\cite{Paranal_atmos}, atmospheric attenuation per airmass $k$ is a sum
of Rayleigh scattering from interactions with atmospheric components
small compared to the wavelength ($k\sim\lambda^{-4}$), Mie scattering
off aerosols of comparable size ($k\sim\lambda^{-1.4}$), cloud
scattering from large water and ice particles ($k\sim\hbox{const}$)
and molecular absorption.  Rayleigh scattering and molecular
absorption normally depend only on the integrated density along the
line of sight and are temporally stable for stable molecule
concentrations (e.g. O$_2$ but not H$_2$O).  Cloud scattering
obviously is extremely variable, particularly over the large field of
view of \PS.  Aerosols arise from volcanic eruptions, smoke, and dust
and are highly variable, both in amplitude as well as spectral shape,
and the $-1.4$ power law is very approximate.  \cite{Paranal_atmos}
make the point that volcanic events should not be thought of as
creating brief increases in aerosol extinction, but instead, times
of low and constant aerosol extinction are exceptionally rare.


In order to manage the complexity of different atmospheric extinction
components as well as to provide the high spectral resolution that can
be important for non-stellar SEDs, we use the MODTRAN program
\citep{MODTRAN} to compute atmospheric transmission to the peak of
Halekala for a range of zenith angles and water vapor content. The
MODTRAN ``Generic Tropical'' model atmosphere was used, with ``Desert
Extinction (Spring-Summer)'' aerosol choice.  No attenuation from
clouds was included.  An alternative atmospheric model from
Atmospheric and Environmental Research\footnote{http://rtweb.aer.com}
has been used by \cite{Paranal_atmos} and we are confident that it
would be equally satisfactory.

For each \PS\ bandpass we integrated a set of power law SEDs against
each of these model atmospheres and created an interpolation function
for the extinction as a function of four variables: $z$ for airmass
($\sec \zeta$ where $\zeta$ is the zenith angle), $h$ for precipitable
water vapor (PWV) (typically 0.65~cm at sea level), $a$ for ``aerosol
exponent'' (nominally 1; we modify the Modtran aerosol component by
applying this power to the aerosol transmission, thereby mostly
affecting the aerosol amplitude), and $p$ for SED power law.  ($p=+2$
for $f_{\nu}\sim\nu^{+2}$ corresponds to an O star with
$(r{-}i)=-0.43$, $p=0$ for $f_{\nu}\sim\hbox{const}$ corresponds to an
F star with $(r{-}i)=0.00$, and $p=-2$ for $f_{\nu}\sim\nu^{-2}$
corresponds to an K5 star with $(r{-}i)=+0.42$.  Note that
$(g{-}r)\sim0.2+1.9(r{-}i)$ in this range.)  The extinction $dm$ in
magnitudes is given by
\begin{equation}
  \ln dm = \ln C +  Z\ln z + A\ln a + P p + \ln h(H_0 + H_1\ln z + H_2\ln h)
\label{eq:modterp}
\end{equation}
The coefficients for each of the \PS\ filters are given in Table~\ref{tab:modterp}.

\begin{table}[htdp]
\caption{\PS\ Extinction Coefficients}
\begin{center}
\begin{tabular}{lrrrrrrrr}
\hline
\hline
  Filter   & $C$   & $Z$   & $A$   & $P$      & $H_0$    & $H_1$    & $H_2$ & err\\
\hline                                                                                            
  \gps     & 0.204 & 0.982 & 0.227 & $ 0.021$ & $ 0.001$ & $-0.000$ & 0.000 & 1.7\\
  \rps     & 0.123 & 0.975 & 0.283 & $ 0.012$ & $ 0.012$ & $-0.000$ & 0.005 & 2.0\\
  \ips     & 0.092 & 0.831 & 0.304 & $ 0.005$ & $ 0.125$ & $-0.011$ & 0.035 & 2.7\\
  \zps     & 0.060 & 0.878 & 0.375 & $-0.004$ & $ 0.330$ & $-0.070$ & 0.055 & 4.9\\
  \yps     & 0.154 & 0.680 & 0.145 & $ 0.014$ & $ 0.549$ & $-0.084$ & 0.024 & 3.5\\
  \wps     & 0.139 & 0.936 & 0.259 & $ 0.075$ & $ 0.029$ & $-0.002$ & 0.009 & 2.2\\
{\it Open} & 0.137 & 0.897 & 0.244 & $ 0.112$ & $ 0.093$ & $-0.018$ & 0.020 & 4.6\\
\hline
\end{tabular}
\end{center}
\tablecomments{The columns contain coefficients described above for
  that interpolate the Modtran extinction calculations each of the
  \PS\ bandpasses.  The final column is the percentage scatter of
  these fits relative to the calculated values.  Note that the
  saturation of molecular lines means that the extinction is {\it not}
  proportional to $\sec\zeta$ ($Z\neq1$), particularly \yps.
  }
\label{tab:modterp}
\end{table}%

The interpolation Formula \ref{eq:modterp} offers only limited
adjustability in the extinction coefficients via the aerosol
transmission exponent $a$, essentially adjusting the aerosol amplitude
but not its spectral shape.  Therefore matching observations of
standards as a function of airmass on a given night may call for
additional term $\delta k\sec\zeta$.  In addition, ozone absorption in
the \rps\ band is significant, and $O_3$ does vary somewhat
(\cite{Paranal_atmos} find a peak-to-peak yearly variation of 0.01 mag
in $k_r$).  The total column of $O_3$ is usually expressed in ``Dobson
units'' (DU, 10~$\mu$m thick layer at STP), and we find the effect of
$DU$ ozone column on the \rps\ extinction coefficient to be $\delta
k_r = 1.0\times10^{-4} (DU-260)$.  (Ozone column can be obtained from
OMI/TOMS satellite measurements\footnote{http://oozoneaq.gsfc.nasa.gov}.)

In order to monitor the water content $h$ of the atmosphere, we
deployed a 180~mm astrograph (the ``spectroscopic sky probe'') with a
coarse diffraction grating across the aperture, and pointed it at the
north celestial pole \citep{polar}.  It has been in continuous
operation since June 2011.  The spectrum of Polaris provides
equivalent widths of water bands, the most important at 723~nm,
822~nm, and 946~nm, as well as the A and B bands of $O_2$.

We found that the atmospheric absorption was accurately matched by the
MODTRAN models, and that we could infer a value for $h$ that is
accurate to about 10\% from the observed equivalent widths.  For
example, MODTRAN models produce an equivalent width for the water band
between 810--836~nm of EW = 0.79~nm~$h^{0.74}\;\sec^{0.75}\zeta$, and
comparison with the Polaris observations allows us to determine $h$.
The mean PWV $h$ of 0.65~cm varies by about 50\% RMS over long
periods, although it tends to be much more stable than that during a
night.  We therefore have adopted PWV of 0.65~cm as the water column
for the nominal \PS\ bandpasses; it affects \ips, \zps, \wps, and
especially \yps.

\subsection{Synthetic Photometry}
\label{sec:synthetic}

We collected the SEDs of 783 spectrophotometric standards, including
59 STIS Calspec photometric standards \citep{Calspec}, which range
from the Sun to Vega to stars fainter than
$V=15$ mag\footnote{http://www.stsci.edu/hst/observatory/cdbs/calspec.html}.
The fundamental basis for this photometry derives from models of
hydrogen white dwarf atmospheres \citep{Bohlin07} and comparisons
between Vega and blackbodies, summarized by \cite{Hayes+Latham75} and
\cite{Hayes85}.
The spectrophotometry of Gunn and Stryker \citep{Gunn+Stryker}, augmented by
Bruzual and Persson to include the UV and
IR provided another 175 SEDs\footnote{http://ftp.stsci.edu/cdbs/grid/bpgs}
There are 379 relatively bright stars from the ``Next Generation
Spectral Library'' from
STScI, although caution is indicated for stars with poor slit
centering\footnote{http://archive.stsci.edu/prepds/stisngsl}.
%
The 4 SDSS spectrophotometric standards from \cite{Fukugita96} were included
as well as their spectrum of Vega.  
The Pickles spectrophotometry library includes 131 stellar SEDs
spanning a range of temperature and luminosity
\citep{Pickles}\footnote{http://cdsarc.u-strasbg.fr/viz-bin/ftp-index?J/PASP/110/863}. 
Finally we included 23 spectrophotometric observations of very cool
stars from the SPEX prism database as well as 11 optical spectra of
brown dwarfs from Mike Cushing (private
communication)\footnote{http://web.mit.edu/ajb/www/browndwarfs/spexprism/index.html}. 

We also assembled Johnson $B$ and $V$ and Cousins $R$ and $I$
bandpasses from \cite{Bessel90} (noting their convention of ``energy
sensitivity functions'' that have units of photons per erg and Vega
normalization). 
The $J$, $H$, and $K_s$ IR bandpasses and zeropoints (``energy
sensitivity'' and Vega normalized) of the 2MASS survey were obtained
from \cite{Cohen03} and the 2MASS
website\footnote{http://www.ipac.caltech.edu/2mass/releases/allsky/doc/sec6\_4a.html}
since 2MASS provides a full-sky homogeneous set of observations.
Note that other definitions of $JHK_s$ such as the ``MKO-NIR'' set
described by \cite{Simons+Tokunaga} or the UKIDSS survey differ somewhat.
%
%
The SDSS bandpasses are presented in \cite{Fukugita96}, but were
derived from the recommendations on the SDSS 
website\footnote{http://www.sdss.org/dr3/instruments/imager/\#filters}.

Finally, we have \PS\ bandpasses that are the product of atmosphere,
optics and detector throughput, and filter.

We multiply all of these SEDs by each bandpass and integrate to obtain
predictions for flux, magnitude, and color (either AB or Vega
depending on the bandpass).  Our calculation keeps careful track of
uncertainties in the SED and tries to estimate uncertainty when an SED
and a filter do not completely overlap.

\subsection{Standard Star Observations}
\label{sec:standards}

MJD~55744 (UT 02 July 2011) was a photometric night during which we
observed a substantial number of spectrophotometric standard stars
from the STIS Calspec \citep{Calspec} tabulation: 1740346, KF01T5,
KF06T2, KF08T3, LDS749B, P177D, and WD1657-343.  These were observed
throughout the night at airmasses between 1 and 2.2 in all six
filters and also with no filter in the beam.  Each observation was
repeated, and exposure times were chosen to stay well clear of any
non-linearities but still permit good accuracy.
In addition, Medium Deep Field 9 (MD09), which overlaps SDSS Stripe82,
was observed a dozen times in each of \gps, \rps, \ips, \zps, and
\yps, providing the opportunity to tie the spectrophotometric data to a
well-observed \PS\ field.  All standard stars were placed on OTA~34 and
cell~33, so their integration was on the same silicon and used the
same amplifier for read-out (gain measured to be 0.97~e$^-$/ADU).

The observations were bias subtracted and flatfielded as part of the
normal IPP processing, and the IPP fluxes (instrumental magnitudes)
were then available for comparison with tabulated SEDs.  The IPP
performs an aperture correction and reports fluxes within a radius of
25 pixels (13\arcsec\ diameter).  

Observations of Polaris on MJD 55744 with the spectroscopic sky probe
had a PWV indistinguishable from the long term mean of 0.65~cm.  

\subsection{Photometry Refinement}
\label{sec:tweaks}

The \PS\ cross section $A(\nu,t)$ for capturing photons is
obtained by multiplying the factors of atmosphere for a given
observation, the in-situ measurements of optics and detector
throughput, and the filter transmission.  In principle there are only
two unknown parameters: a single overall normalization factor,
required because the in-situ throughput measurements did not attempt
to evaluate the net collecting area of the telescope, and the aerosol
extinction exponent $a$ for the night of the standard star
observations.

In practice, we found that small ``tweaks'' were required to bring
observations into agreement with spectrophotometry.  The need for
these tweaks is not surprising because our measurement technique
currently has the potential for systematic error at the several
percent level (for example, we sample the telescope pupil at only one
point, ghost image and scattered light compensation, chromatic effects
from fiber in illumination of photodiode, etc), and we are trying to
achieve 1\% accuracy.  However, the excellent agreement between the
laser and Barr measurements of the filter band edges and transmission
wiggles led us to parameterize the tweaks as a smooth adjustment to
the throughput function and individual transmission adjustments for each
filter\footnote{We emphasize that we are {\it not} attempting to determine
  zeropoints for each filter individually; we determine {\it one}
  zeropoint for the \PS\ system and these transmission offsets and
  throughput tweaks represent the extent to which we were unsuccessful
  (3\%) in our in-situ measurements (or conceivably error in the
  spectrophotometric standard SEDs).}.  There is an ambiguity between
whether tweaks should be applied to throughput or filter, and we have
attempted to disentangle them as best we can using the information
from overlapping bandpasses (\wps\ overlaps \gps, \rps, and \ips) and
standard star observations with no filter.

We adjust a total of 12 parameters for the \PS\ system: 9 parameters
provide offset and spectral tilt tweaks for throughput and each filter
(expected to be durable at the 1\% level for very long periods), 2
parameters characterize the aerosol extinction (changes nightly), and
1 parameter sets the overall collecting area (expected to slowly
change with dust and degradation of optical surfaces).

The \PS\ no-filter cross-section $A_0(\nu)$ consists of the area of a
1.8~m disk, times the geometrical loss from secondary and baffles of
0.62 derived from ray tracing, times in-situ throughput
measurements, adjusted for IR light scattering in the Si and
normalized to a peak of 0.70 (the peak of the product of Al
reflectivities, AR coatings, and CCD QE), times the tweak function.
The tweak function we adopted consists of a natural spline with five
knots at 400, 550, 700, 850, and 1000~nm and values we determined to
be $0.035$, $0.113$, $0.113$, $0.081$, and $0.022$ mag (positive
meaning less sensitive).  The mean across the optical of 0.085 mag
simply measures the wavelength-independent deviation from the
arbitrary 0.70 peak throughput and amounts to a normalization
correction.  The spectral variation of 0.030 mag RMS is the mis-match
between our in-situ instrumental throughput measurements and the
spectrophotometric standard observations, after making the aerosol
adjustment to the atmospheric transmission.  This tweak function is
illustrated in Fig~\ref{fig:ps1trans}.

The filter-specific tweaks were determined to be $0.012$, $0.019$,
$0.009$, $-0.009$, $-0.010$, $-0.005$ mag for \gps, \rps, \ips, \zps,
\yps, and \wps\ (positive is less sensitive; the mean of \ips\ and
\zps\ is constrained to zero).

The procedure for determining these parameters involves iterating a
comparison between synthetic photometry using spectrophotometric SEDs
with observations of standard stars and stellar locus. The combination
of our atmospheric transmission model and the system transmission
measurements produce (untweaked) synthetic photometry that disagrees
with the observations by 0.1 mag peak-to-peak from \gps\ to \yps.
We have elected to trust the Calspec SEDs as the foundational
calibration data, and we adjust the response functions to achieve
photometric consistency.

For each of the seven Calspec spectrophotometric standards observed on
MJD~55744 we calculated predictions for the flux (including color
terms appropriate for the actual filter location and OTA on which they
were observed), and adjusted the parameters to match the observations.
We found that the variation with airmass called for modification of
the MODTRAN extinction with an aerosol exponent $a=0.7$ and an
additional $\delta k=-0.02$~mag/airmass (i.e. aerosols were lighter
than the MODTRAN default by about 30\% and had a steeper rise at bluer
wavelengths).  The standards had a large enough diversity in color
($-0.38<(r-i)<+0.35$) to provide some constraint on the filter tilt
parameters (spline knots).

\begin{figure}[htbp]
\begin{center}
\centerline{\includegraphics[width=\figwid]{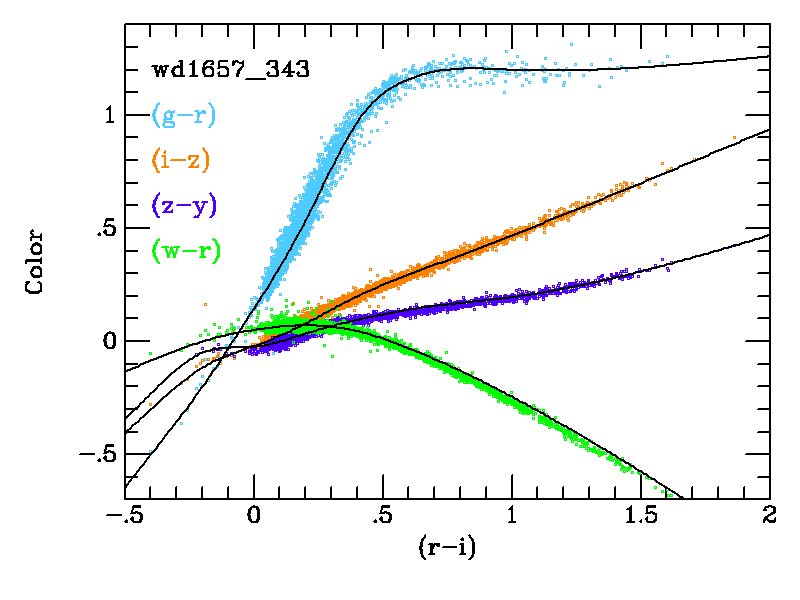}}
\caption{The stellar locus calculated from SED integration is
  plotted over the locus of the stars near WD1657-343.  This
  field has the lowest Galactic extinction of the standards, and has
  the longest integration time.  The stellar magnitudes were averaged
  from all the observations.}
\label{fig:locuswd}
\end{center}
\end{figure}

As another check, we computed a ``stellar locus'' from all of the
spectrophotometric standards.  This involves de-reddening the SEDs of
galactic extinction, computing synthetic colors in the
\PS\ bandpasses, and fitting various colors as a function of
$(r{-}i)_{P1}$.  Uncertainties in Galactic extinction were propagated
into the colors.  Each of the standard star observations and MD09
includes thousands of stars over the field of view, and these
magnitudes were de-reddened as well using \cite{SFD} (SFD) values for
Galactic extinction.  The comparison provides us with a second
constraint on the tweak parameters, and is the reason that the mean
offsets of the standard star observations are not simply zero.  The
huge color range of field stars creates the strongest constraint on
the filter tilt parameters.  Figure~\ref{fig:locuswd} shows the
observed stellar colors with the spline curves from the
spectrophotometric standards overplotted.

We also calculated \PS-SDSS color transformations, computed
\PS\ magnitudes from SDSS magnitudes in Stripe82 obtained from Zeljko
Ivezic\footnote{http://www.astro.washington.edu/users/ivezic/sdss/catalogs},
and compared them to the observed magnitudes of stars in the MD09
observations on MJD~55744.  This was {\it not} used to adjust
parameters, however.  Table~\ref{tab:magtest} shows the difference
between the fluxes observed for the spectrophotometric standard
stars and the SDSS stars in MD09 and magnitudes calculated from SED and SDSS
magnitudes transformed to the \PS\ system.
\begin{table}[htdp]
\caption{\PS\ Photometric Consistency Checks.}
\begin{center}
\begin{tabular}{lrrrrr}
\hline
\hline
 Filter& Std    & $\pm$ &  SDSS  & $\pm$ & N\\
\hline
  \gps & $-$0.004 & 0.007 &    0.014 & 0.012 & 2644\\
  \rps & $-$0.005 & 0.006 & $-$0.019 & 0.010 & 3072\\
  \ips &    0.008 & 0.009 &    0.008 & 0.011 & 2850\\
  \zps & $-$0.009 & 0.007 &    0.015 & 0.011 & 2816\\
  \yps &    0.005 & 0.010 &    0.001 & 0.013 & 2150\\
  \wps &    0.002 & 0.011 &     ---  &  ---  &  ---\\
\hline
\end{tabular}
\end{center}
\tablecomments{The columns are the filter, average difference for the
  standard stars between observed instrumental magnitude (flux) and that
  predicted from SED, scatter among the $\sim24$ observations, average
  difference between \PS\ magnitude and SDSS magnitude, RMS scatter,
  and number of stars compared.  The SDSS comparison is restricted to
  stars in a 3 magnitude range: $15<g<18$ to $13<y<16$.}
\label{tab:magtest}
\end{table}%

\section{THE \PS\ PHOTOMETRIC SYSTEM}
\label{sec:psphot}

\begin{figure}[htbp]
\begin{center}
\centerline{\includegraphics[width=\figwid]{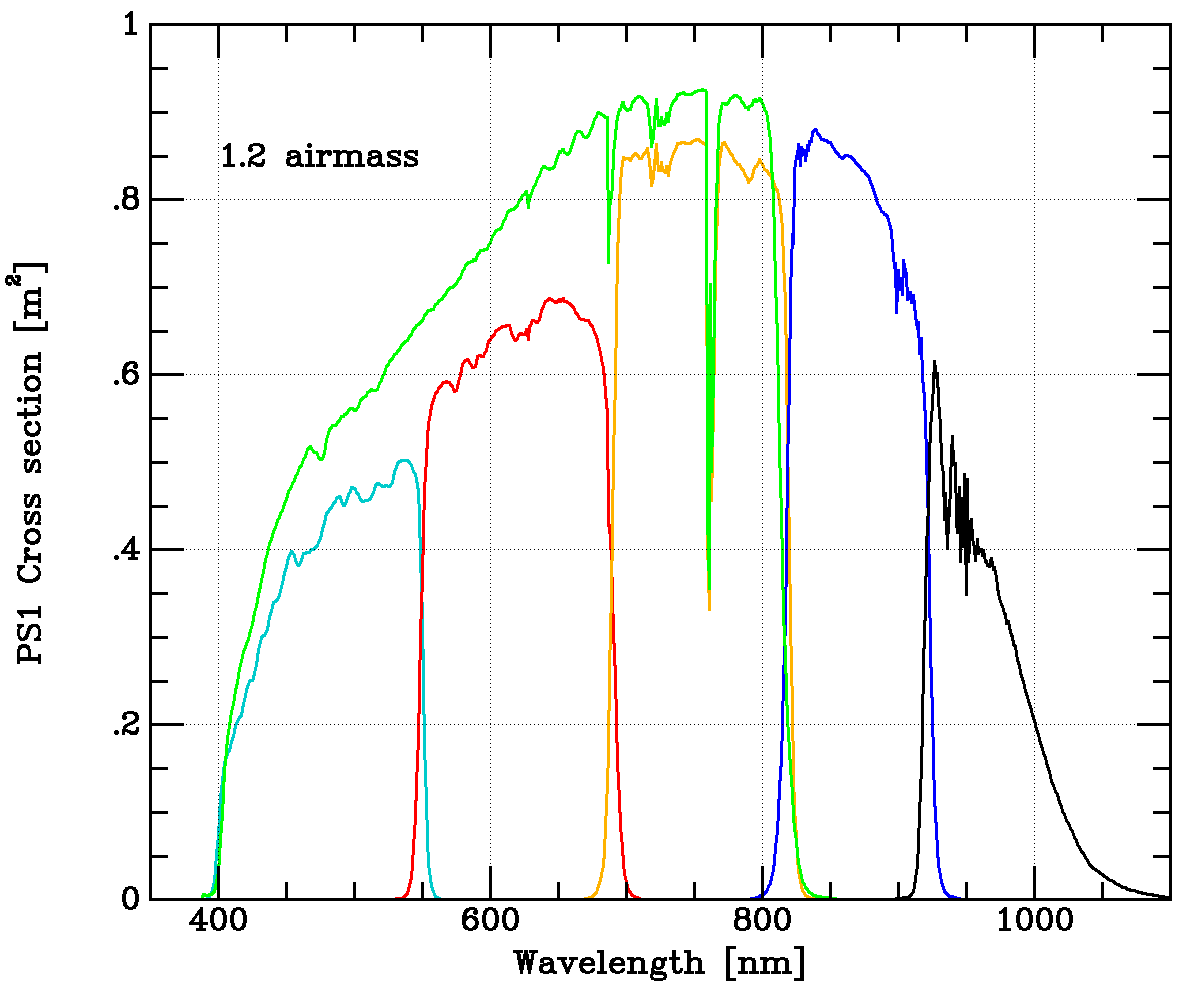}}
\caption{The \PS\ capture cross section $A(\nu)$ in m$^2$-e$^-$/photon
  to produce a detected $e^-$ for an incident photon
  for the six \PS\ bandpasses.  This is at the standard airmass of
  1.2, with standard PWV of 0.65~cm and aerosol exponent 0.7.  Summary
  properties of each bandpass are found in Table~\ref{tab:pszp}.}
\label{fig:ps1xsec}
\end{center}
\end{figure}

After iteration to determine the best fit parameters, we present
Figure~\ref{fig:ps1xsec} showing the net \PS\ collecting area as a
function of wavelength for the six filters, i.e. $A(\nu)$.  This is
the product of the MODTRAN atmosphere at 1.2 airmass from elevation
3~km with 0.65~mm of PWV at sea level and 0.7 aerosol, the vignetted
collecting area, the throughput function of Figure~\ref{fig:ps1trans},
and the filter transmissions.

A detailed spectral tabulation of the \PS\ bandpasses is found in
Table~\ref{tab:bandpass}.
\begin{table}[htdp]
\caption{\PS\ Bandpasses}
\begin{center}
\begin{tabular}{rrrrrrrrrrr}
\hline
\hline
$\lambda$ & {\it Open} & \gps & \rps & \ips & \zps & \yps & \wps & Aero & Ray & Mol\\
\hline
\ldots&\ldots&\ldots&\ldots&\ldots&\ldots&\ldots&\ldots&\ldots&\ldots&\ldots\\
 550 & 0.675 & 0.323 & 0.367 & 0.000 & 0.000 & 0.000 & 0.662 & 0.962 & 0.924 & 0.971\\
 551 & 0.678 & 0.248 & 0.423 & 0.000 & 0.000 & 0.000 & 0.665 & 0.962 & 0.924 & 0.971\\
 552 & 0.682 & 0.175 & 0.469 & 0.000 & 0.000 & 0.000 & 0.668 & 0.962 & 0.925 & 0.971\\
 553 & 0.685 & 0.113 & 0.504 & 0.000 & 0.000 & 0.000 & 0.670 & 0.962 & 0.925 & 0.971\\
 554 & 0.688 & 0.068 & 0.528 & 0.000 & 0.000 & 0.000 & 0.672 & 0.963 & 0.926 & 0.970\\
 555 & 0.689 & 0.041 & 0.544 & 0.000 & 0.000 & 0.000 & 0.673 & 0.963 & 0.927 & 0.970\\
 556 & 0.690 & 0.025 & 0.556 & 0.000 & 0.000 & 0.000 & 0.674 & 0.963 & 0.927 & 0.969\\
 557 & 0.691 & 0.015 & 0.565 & 0.000 & 0.000 & 0.000 & 0.674 & 0.963 & 0.928 & 0.968\\
 558 & 0.691 & 0.009 & 0.570 & 0.000 & 0.000 & 0.000 & 0.675 & 0.963 & 0.928 & 0.968\\
 559 & 0.693 & 0.006 & 0.575 & 0.000 & 0.000 & 0.000 & 0.678 & 0.963 & 0.929 & 0.967\\
 560 & 0.695 & 0.003 & 0.578 & 0.000 & 0.000 & 0.000 & 0.680 & 0.963 & 0.929 & 0.966\\
\ldots&\ldots&\ldots&\ldots&\ldots&\ldots&\ldots&\ldots&\ldots&\ldots&\ldots\\
\hline
\end{tabular}
\end{center}
\tablecomments{The columns are wavelength [nm] and capture
  cross-section [m$^2$-e$^-$/photon] for each of the \PS\ bandpasses,
  {\it including} the nominal 1.2 airmasses of atmospheric extinction.
  The last three columns list the transmission of the \PS\ standard
  atmosphere from aerosol scattering, Rayleigh scattering, and
  molecular absorption.
  Table~\ref{tab:bandpass} is published in its entirety in the
  electronic edition of the Astrophysical Journal.  A portion is shown
  here for guidance regarding its form and content.}
\label{tab:bandpass}
\end{table}%
Summary parameters of the \PS\ bandpasses are found in Table~\ref{tab:pszp}.  
The ``zeropoints'' are the AB magnitude of a
neutral color (constant $f_\nu$) star that would produce 1~e$^-$/sec
in the detector with 1.2 airmasses of extinction.  We also list the
net atmospheric extinction {\it at 1.2 airmasses} we expect to see for
an SED of constant $f_\nu$, so the sum of these two numbers is the
``top of atmosphere'' zeropoint for the \PS\ system.  We list these
separately to emphasize that, however important it may be,
extrapolation to ``top of atmosphere'' depends on SED of source and
there is no unique correct answer, and also that extinction is not
linear in airmass.  Equation~\ref{eq:modterp} can be used to explore
these dependencies.  The sky brightnesses are AB mag
per square arcsec, calculated for a dark sky model and observed
between 2010-11-12 and 2011-05-12.  Most of the discrepancy between
calculated and observed comes from the degree to which moonlight
impinges on normal operations, although the \yps\ brightness has 
recently been reduced by introduction of a new baffle.
\begin{table}[htdp]
\caption{\PS\ Bandpass Parameters}
\begin{center}
\begin{tabular}{lcrrrccccc}
\hline
\hline
  Filter & $\langle A\rangle$ & $\lambda_{eff}$ & $\lambda_B$ & $\lambda_R$ & 
  ZP & Extinct & $\mu$ & $\mu_{obs}$\\ 
\hline
  \gps     & 0.1212 & 481 & 414 &  551 & 24.56 & 0.22 & 22.12 & 21.92\\
  \rps     & 0.1463 & 617 & 550 &  689 & 24.76 & 0.13 & 20.97 & 20.83\\
  \ips     & 0.1435 & 752 & 690 &  819 & 24.74 & 0.09 & 20.18 & 19.79\\
  \zps     & 0.0980 & 866 & 818 &  922 & 24.33 & 0.05 & 19.27 & 19.24\\
  \yps     & 0.0393 & 962 & 918 & 1001 & 23.33 & 0.13 & 18.43 & 18.24\\
  \wps     & 0.4739 & 608 & 433 &  815 & 26.04 & 0.15 & 20.86 & 20.62\\
{\it Open} & 0.6463 & 655 & 431 &  971 & 26.37 & 0.14 & 20.12 & 20.00\\
\hline
\end{tabular}
\end{center}
\tablecomments{The columns are the filter, ``net cross section''
  [m$^2$] for $f_\nu$=const through this filter at 1.2 airmasses
  ($\int A(\nu)d\ln\nu$), filter ``pivot'' wavelength [nm] described
  by \cite{Bessel12}
  ($\int\lambda A(\nu)d\ln\nu/\langle A\rangle$), bandpass blue and
  red wavelengths [nm] obtained from a least-squares fit of a square
  bandpass, zeropoint at 1.2 airmasses [AB mag], extinction at 1.2
  airmasses [mag] ({\it not} extinction per airmass!), calculated dark
  sky brightness [mag/\arcsec], and median observed sky brightness
  [mag/\arcsec].}
\label{tab:pszp}
\end{table}%

\subsection{The \PS\ Stellar Locus}

The derivation of the synthetic \PS\ stellar locus from the library of
SEDs mentioned above first required removal of Galactic reddening.  We
started by undoing the correction applied by \cite{Gunn+Stryker} for
Galactic extinction, returning them to ``as-observed'' SEDs, although
we kept their estimates of $A_V$.  We then estimated a $V$ band
extinction value for the rest of the stars by using the
\cite{Parenago} model recommended by \cite{Groenewegen08} (scale
height of 90~pc and visual extinction of 1.08~mag/kpc), and using
parallaxes from
SIMBAD\footnote{http://http://simbad.u-strasbg.fr/simbad/}.
Uncertainties in parallax were folded into flux and color
uncertainties.  Given a value for $A_V$ and adopting $R_V=3.1$, we
used the extinction curves from \cite{Fitzpatrick99} to calculate
stellar SEDs with no reddening from dust, and then integrated
magnitudes in all bandpasses

Table~\ref{tab:locus} lists spline knots fitted to this locus, using
$(r{-}i)$ as the independent variable.  The synthetic \PS\ colors are
seen in Figure~\ref{fig:locuswd}.  The prominent wiggle visible in
the $(z{-}y)$ locus at $(r{-}i)\sim0$, also visible in $(i{-}z)$,
arises from Paschen absorption that peaks at spectral type A.
\begin{table}[htdp]
\caption{\PS\ Synthetic Stellar Locus}
\begin{center}
\begin{tabular}{rrrrrrrrr}
\hline
\hline
$(r{-}i)$&$(g{-}r)$&$(i{-}z)$&$(z{-}y)$&$(z{-}J)$&$(z{-}H)$&$(y{-}J)$&$(w{-}r)$&$(O{-}r)$\\
\hline
 $-0.4$&$-0.50$&$-0.290$&$-0.210$&$0.12$&$0.05$&$0.34$&$-0.085$&$ 0.015$\\
 $-0.2$&$-0.19$&$-0.110$&$-0.050$&$0.48$&$0.50$&$0.50$&$ 0.000$&$ 0.070$\\
 $ 0.0$&$ 0.15$&$-0.030$&$-0.025$&$0.70$&$0.87$&$0.70$&$ 0.050$&$ 0.060$\\
 $ 0.2$&$ 0.55$&$ 0.090$&$ 0.035$&$0.89$&$1.28$&$0.86$&$ 0.070$&$-0.010$\\
 $ 0.4$&$ 0.97$&$ 0.200$&$ 0.095$&$1.14$&$1.82$&$1.00$&$ 0.045$&$-0.120$\\
 $ 0.6$&$ 1.16$&$ 0.295$&$ 0.140$&$1.22$&$1.96$&$1.11$&$-0.030$&$-0.280$\\
 $ 1.0$&$ 1.20$&$ 0.470$&$ 0.195$&$1.31$&$2.00$&$1.10$&$-0.245$&$-0.670$\\
 $ 2.0$&$ 1.26$&$ 0.940$&$ 0.470$&$1.23$&$2.12$&$0.87$&$-0.940$&$-1.820$\\
\hline
\end{tabular}
\end{center}
\tablecomments{The columns provide knots for a natural spline for 
  various \PS\ and \PS-2MASS colors as a function of $(r{-}i)_{\rm P1})$.}
\label{tab:locus}
\end{table}%

\begin{figure}[htbp]
\begin{center}
\centerline{\includegraphics[width=\figwid]{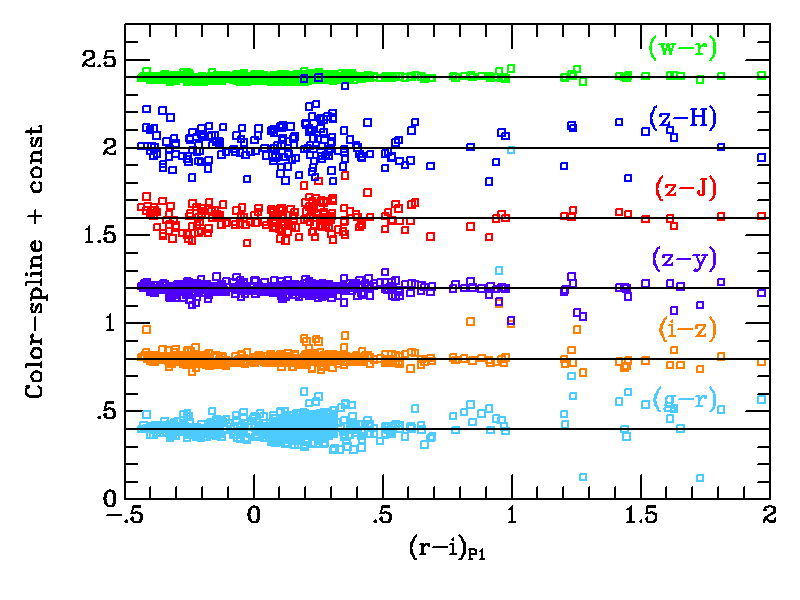}}
\caption{The residual of the \PS\ colors we calculate for the SEDs
  relative to the spline fits.  The RMS of
  the residual scatter is 0.09, 0.03, 0.03, 0.01, 0.06, 0.10 mag for 
  $(g{-}r)$, $(i{-}z)$, $(z{-}y)$, $(w{-}r)$, $(z{-}J)$, and $(z{-}H)$.}
\label{fig:ps1locresid}
\end{center}
\end{figure}

The residuals of the synthetic colors from the 783 SEDs relative to
the spline fits in Figure~\ref{fig:ps1locresid} demonstrate that
the splines have accurately captured the variation.

\subsection{Stellar Color Transformations}

\begin{figure}[ht]
\begin{center}$
\begin{array}{ccc}
\includegraphics[width=3in]{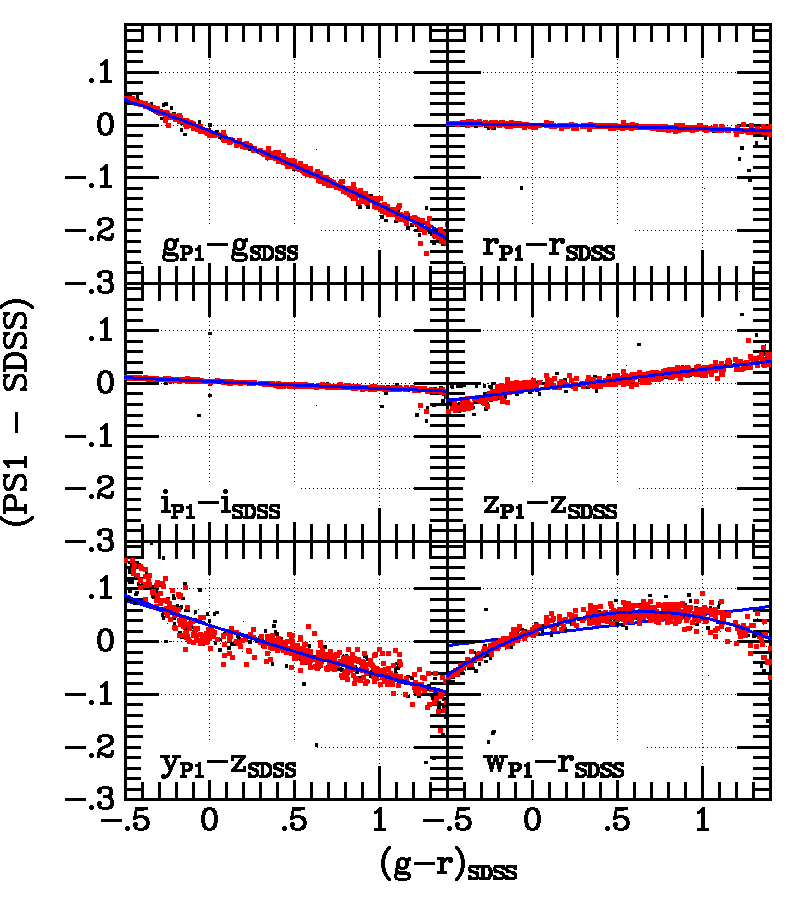} &
\includegraphics[width=3in]{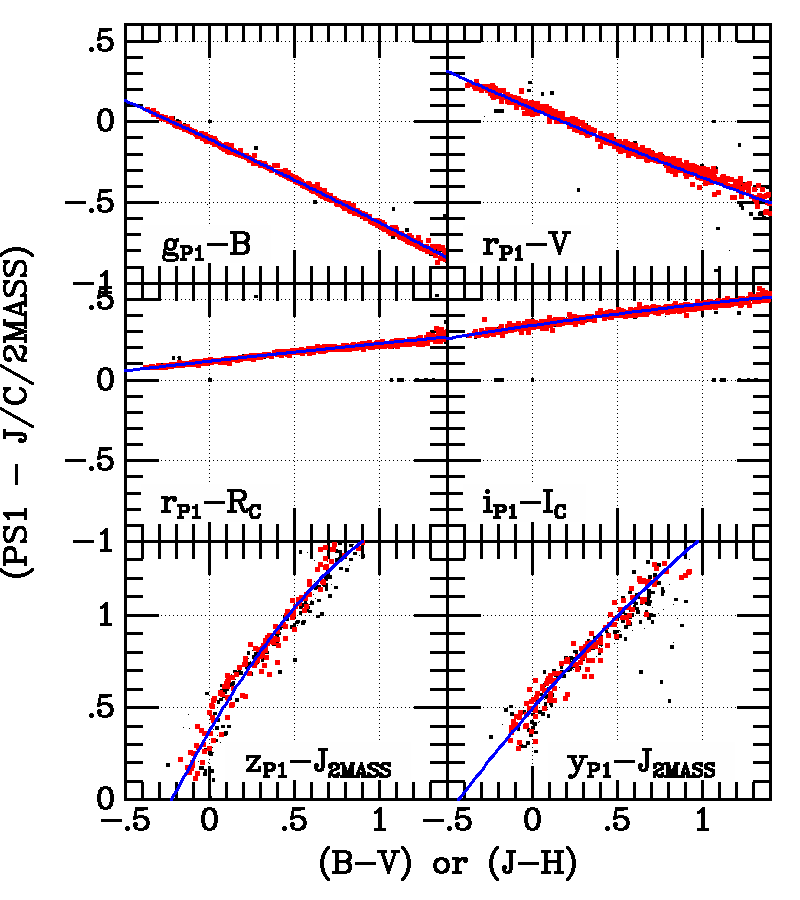}
\end{array}$
\end{center}
\caption{Comparison between the \PS\ and SDSS bandpasses as a function
  of SDSS color (left) and Johnson, Cousins, and
  2MASS bandpasses as a function $(B{-}V)$ or $(J{-}H)$ (right).}
\label{fig:psfrom}
\end{figure}

\begin{figure}[ht]
\begin{center}$
\begin{array}{ccc}
\includegraphics[width=3in]{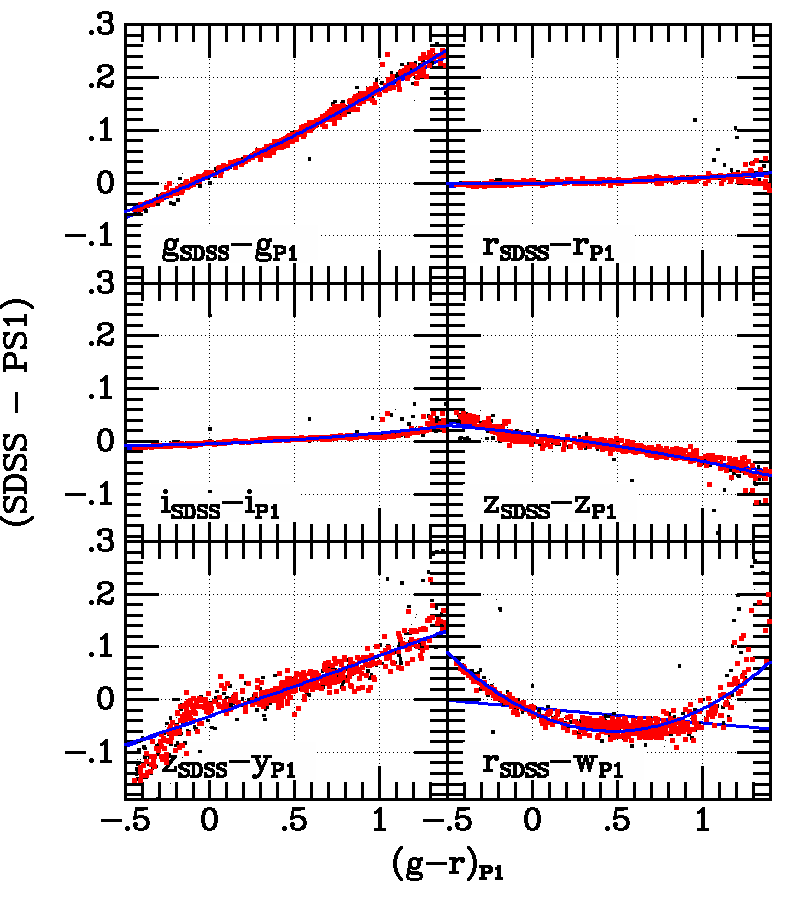} &
\includegraphics[width=3in]{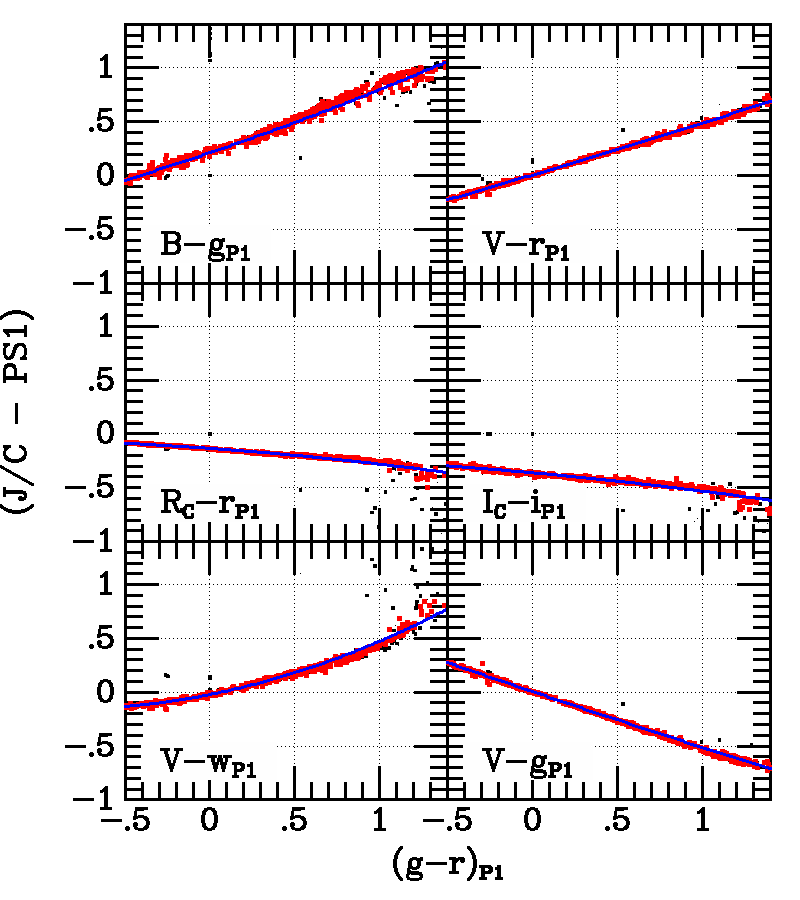}
\end{array}$
\end{center}
\caption{Comparison between the SDSS and \PS\ bandpasses as a function
  of $(g{-}r)_{P1}$ (left) and Johnson and Cousins
  versus \PS\ as a function $(g{-}r)_{P1}$ (right).}
\label{fig:psto}
\end{figure}

We used the synthetic magnitudes from the SEDs to fit for conversions
between the \PS\ photometric system and SDSS, Johnson/Cousins
(Vega), and 2MASS (Vega).  Both linear and quadratic versions are
provided, with coefficients
\begin{equation}
y = A_0 + A_1 x + A_2 x^2 = B_0 + B_1 x.
\label{eq:xform}
\end{equation}
Figures \ref{fig:psfrom} and \ref{fig:psto} illustrate these
relationships, and Table~\ref{tab:xform} provides the coefficients.
We stress that these are computed for stellar SEDs and use for other
SEDs may be less accurate.  The marked deviations of \yps\ and
\zps\ relative to $z_{SDSS}$ arise because of Paschen absorption
and will differ for blue objects that lack hydrogen lines.
The figures provide guidance about the validity of the linear or quadratic
fits.
\begin{table}[htdp]
\caption{\PS\ Bandpass Transformations}
\begin{center}
\begin{tabular}{ccrrrrrrr}
\hline
\hline
$x$ & $y$ & $A_0$ & $A_1$ & $A_2$ & $\pm$ & $B_0$ & $B_1$ & $\pm$ \\
\hline
$(g{-}r)_{SDSS}$&$(g_{P1}{-}g_{SDSS})$&$-0.011$&$-0.125$&$-0.015$&$0.006$&$-0.012$&$-0.139$&$0.007$\\
$(g{-}r)_{SDSS}$&$(r_{P1}{-}r_{SDSS})$&$ 0.001$&$-0.006$&$-0.002$&$0.002$&$ 0.000$&$-0.007$&$0.002$\\
$(g{-}r)_{SDSS}$&$(i_{P1}{-}i_{SDSS})$&$ 0.004$&$-0.014$&$ 0.001$&$0.003$&$ 0.004$&$-0.014$&$0.003$\\
$(g{-}r)_{SDSS}$&$(z_{P1}{-}z_{SDSS})$&$-0.013$&$ 0.040$&$-0.001$&$0.009$&$-0.013$&$ 0.039$&$0.009$\\
$(g{-}r)_{SDSS}$&$(y_{P1}{-}z_{SDSS})$&$ 0.031$&$-0.106$&$ 0.011$&$0.023$&$ 0.031$&$-0.095$&$0.024$\\
$(g{-}r)_{SDSS}$&$(w_{P1}{-}r_{SDSS})$&$ 0.018$&$ 0.118$&$-0.091$&$0.012$&$ 0.012$&$ 0.039$&$0.025$\\
\hline
$(B{-}V)$&$(g_{P1}{-}B)$              &$-0.108$&$-0.485$&$-0.032$&$0.011$&$-0.104$&$-0.523$&$0.013$\\
$(B{-}V)$&$(r_{P1}{-}V)$              &$ 0.082$&$-0.462$&$ 0.041$&$0.025$&$ 0.077$&$-0.415$&$0.025$\\
$(B{-}V)$&$(r_{P1}{-}R_{C})$         &$ 0.117$&$ 0.128$&$-0.019$&$0.008$&$ 0.119$&$ 0.107$&$0.009$\\
$(B{-}V)$&$(i_{P1}{-}I_{C})$         &$ 0.341$&$ 0.154$&$-0.025$&$0.012$&$ 0.343$&$ 0.126$&$0.013$\\
$(J_{2MASS}{-}H_{2MASS})$&$(z_{P1}{-}J_{2MASS})$&$ 0.418$&$ 1.594$&$-0.603$&$0.068$&$ 0.428$&$ 1.260$&$0.073$\\
$(J_{2MASS}{-}H_{2MASS})$&$(y_{P1}{-}J_{2MASS})$&$ 0.528$&$ 0.962$&$-0.069$&$0.061$&$ 0.531$&$ 0.916$&$0.061$\\
\hline
$(g{-}r)_{P1}$&$(g_{SDSS}{-}g_{P1})$  &$ 0.013$&$ 0.145$&$ 0.019$&$0.008$&$ 0.014$&$ 0.162$&$0.009$\\
$(g{-}r)_{P1}$&$(r_{SDSS}{-}r_{P1})$  &$-0.001$&$ 0.004$&$ 0.007$&$0.004$&$-0.001$&$ 0.011$&$0.004$\\
$(g{-}r)_{P1}$&$(i_{SDSS}{-}i_{P1})$  &$-0.005$&$ 0.011$&$ 0.010$&$0.004$&$-0.004$&$ 0.020$&$0.005$\\
$(g{-}r)_{P1}$&$(z_{SDSS}{-}z_{P1})$  &$ 0.013$&$-0.039$&$-0.012$&$0.010$&$ 0.013$&$-0.050$&$0.010$\\
$(g{-}r)_{P1}$&$(z_{SDSS}{-}y_{P1})$  &$-0.031$&$ 0.111$&$ 0.004$&$0.024$&$-0.031$&$ 0.115$&$0.024$\\
$(g{-}r)_{P1}$&$(r_{SDSS}{-}w_{P1})$  &$-0.024$&$-0.149$&$ 0.155$&$0.018$&$-0.016$&$-0.029$&$0.031$\\
\hline
$(g{-}r)_{P1}$&$(B{-}g_{P1})$         &$ 0.212$&$ 0.556$&$ 0.034$&$0.032$&$ 0.213$&$ 0.587$&$0.034$\\
$(g{-}r)_{P1}$&$(V{-}r_{P1})$         &$ 0.005$&$ 0.462$&$ 0.013$&$0.012$&$ 0.006$&$ 0.474$&$0.012$\\
$(g{-}r)_{P1}$&$(R_{C}{-}r_{P1})$    &$-0.137$&$-0.108$&$-0.029$&$0.015$&$-0.138$&$-0.131$&$0.015$\\
$(g{-}r)_{P1}$&$(I_{C}{-}i_{P1})$    &$-0.366$&$-0.136$&$-0.018$&$0.017$&$-0.367$&$-0.149$&$0.016$\\
$(g{-}r)_{P1}$&$(V{-}w_{P1})$         &$-0.021$&$ 0.299$&$ 0.187$&$0.025$&$-0.011$&$ 0.439$&$0.035$\\
$(g{-}r)_{P1}$&$(V{-}g_{P1})$         &$ 0.005$&$-0.536$&$ 0.011$&$0.012$&$ 0.006$&$-0.525$&$0.012$\\
\hline
\end{tabular}
\end{center}
\tablecomments{The table provides the coefficients for Equation~\ref{eq:xform}.}
\label{tab:xform}
\end{table}%

\subsection{\PS\ Galactic Extinction}

\begin{figure}[htbp]
\begin{center}
\centerline{\includegraphics[width=\figwid]{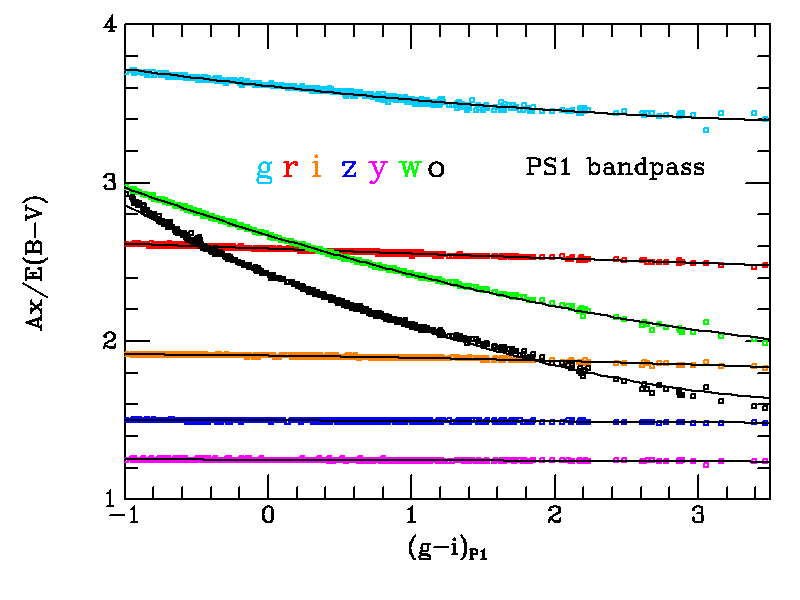}}
\caption{Computed galactic extinction coefficient, $A_x/E(B{-}V)$ in
  the \PS\ bandpasses as a function of stellar color. (Note that
  $E(B{-}V)$ from the SFD catalog should be multiplied by 0.88.)}
\label{fig:galext}
\end{center}
\end{figure}

Equipped with the \PS\ bandpasses, we calculate the
effects of Galactic extinction by applying 0.1 mag of $E(B{-}V)$
Galactic extinction to each of the SEDs and fitting the dimming in
each of the \PS\ bandpasses as a function of unreddened, \PS\ stellar
color.  The extinction curve is from \cite{Fitzpatrick99} using
$R_V=3.1$, and the fits are valid for $-1<(g-i)<4$.  These curves are
illustrated in Figure~\ref{fig:galext}.  ($(g{-}i)\sim0.2+2.9(r{-}i)$
for $(r{-}i)<0.5$.)
\begin{align}
  Ag/E(B{-}V) &= 3.613 - 0.0972 (g{-}i) + 0.0100 (g{-}i)^2\\
  Ar/E(B{-}V) &= 2.585 - 0.0315 (g{-}i)                   \\
  Ai/E(B{-}V) &= 1.908 - 0.0152 (g{-}i)                   \\
  Az/E(B{-}V) &= 1.499 - 0.0023 (g{-}i)                   \\
  Ay/E(B{-}V) &= 1.251 - 0.0027 (g{-}i)                   \\
  Aw/E(B{-}V) &= 2.672 - 0.2741 (g{-}i) + 0.0247 (g{-}i)^2\\
  Ao/E(B{-}V) &= 2.436 - 0.3816 (g{-}i) + 0.0441 (g{-}i)^2
\end{align}
Note that \cite{Schlafly11} recommend a recalibration of the $E(B{-}V)$ from
\cite{SFD} which amounts to multiplication by 0.88.  Therefore when
the formulae above are multiplied by $E(B{-}V)$ in order to obtain
a \PS\ extinction, they should also be multiplied by an additional
factor of 0.88 if $E(B{-}V)$ is derived from SFD.

\subsection{Filter and Detector Color Terms}

The \PS\ filter's response varies as a function of field angle,
although we believe them to be quite uniform as a function of azimuth.
As a function of angle off of the boresight we list color terms for
stellar SEDs in Table~\ref{tab:filtclr}, meaning the slope of the
response in each filter as a function of $(r{-}i)$.  (This creates
offsets in response to SEDs of different color than the color of the
flatfields, which is approximately that of a K star.)  The units are
magnitude per unit $(r{-}i)$ with the usual sign: negative implies
more sensitivity for redder SEDs.  The \gps\ filter in particular is
more red sensitive at large field angle because the red edge of the
bandpass shifts to the red by almost 10~nm.  These offsets do not
change for SEDs redder than $(r{-}i)=0.5$.
\begin{table}[htdp]
\caption{\PS\ Filter Color Terms}
\begin{center}
\begin{tabular}{rrrrrrr}
\hline
\hline
$\theta$&$g^\prime-g$&$r^\prime-r$&$i^\prime-i$&$z^\prime-z$&$y^\prime-y$&$w^\prime-w$\\
\hline
$0.00$&$-0.008$&$-0.005$&$-0.006$&$-0.005$&$-0.011$&$-0.009$\\
$0.15$&$-0.006$&$-0.003$&$-0.006$&$-0.005$&$-0.011$&$-0.009$\\
$0.30$&$ 0.002$&$-0.000$&$-0.002$&$-0.003$&$-0.010$&$-0.008$\\
$0.45$&$ 0.002$&$ 0.002$&$-0.003$&$-0.002$&$-0.008$&$-0.007$\\
$0.60$&$ 0.002$&$ 0.003$&$-0.004$&$-0.003$&$-0.006$&$-0.004$\\
$0.75$&$ 0.003$&$ 0.004$&$-0.001$&$-0.002$&$-0.004$&$ 0.002$\\
$0.90$&$ 0.002$&$ 0.004$&$ 0.001$&$-0.000$&$-0.001$&$ 0.005$\\
$1.05$&$-0.003$&$ 0.002$&$-0.000$&$-0.001$&$ 0.001$&$ 0.006$\\
$1.20$&$-0.006$&$ 0.000$&$ 0.002$&$-0.001$&$ 0.003$&$ 0.008$\\
$1.35$&$-0.010$&$ 0.001$&$ 0.005$&$ 0.002$&$ 0.005$&$ 0.008$\\
$1.50$&$-0.020$&$-0.002$&$ 0.007$&$ 0.006$&$ 0.005$&$ 0.005$\\
$1.65$&$-0.033$&$-0.007$&$ 0.009$&$ 0.010$&$ 0.004$&$ 0.005$\\
\hline
\end{tabular}
\end{center}
\tablecomments{The table provides color terms [mag/mag$(r{-}i)$] for
  each filter as a function of field angle [deg].
  These offsets do not change for SEDs redder than $(r{-}i)=0.5$.}
\label{tab:filtclr}
\end{table}%

The bandpass shapes have negligible sensitivity to CCD temperature
except in the \yps\ band, where the CCDs become more sensitive by
$-0.0004$~mag/K/$(r{-}i)$.  Note that this is the {\it differential}
sensitivity as a function of color --- the overall sensitivity
increase is about an order of magnitude greater, $\sim-0.003$~mag/K.

There is some variation in QE between OTAs, but the color sensitivity
is small in \rps, \ips, \zps, and \yps (less than $0.01$mag/mag).
Variations in the AR coatings do create sensitivity changes in
\gps\ and \wps, however.  Table~\ref{tab:otaclr} lists the \gps\ color
terms for each OTA.  To be explicit, OTA~34 has a color term of $+0.020$,
meaning that it has 1\% greater response than the mean of all OTAs
for an SED of $(r{-}i)=0$ than it does for $(r{-}i)=0.5$.
\begin{table}[htdp]
\caption{\PS\ OTA Color Terms for \gps}
\begin{center}
\begin{tabular}{rrrrrrrr}
\hline
\hline
   OTA77&$-0.018$&$-0.007$&$-0.001$&$-0.042$&$-0.012$&$-0.012$&OTA70\\
$-0.045$&$ 0.015$&$-0.007$&$-0.002$&$ 0.017$&$-0.003$&$-0.016$&$-0.049$\\
$-0.012$&$-0.001$&$ 0.007$&$ 0.008$&$-0.012$&$ 0.005$&$ 0.037$&$-0.005$\\
$-0.016$&$ 0.003$&$ 0.007$&$-0.022$&$ 0.020$&$ 0.038$&$ 0.003$&$-0.017$\\
$ 0.001$&$-0.012$&$-0.002$&$-0.025$&$ 0.003$&$ 0.001$&$ 0.006$&$-0.004$\\
$-0.008$&$-0.003$&$ 0.000$&$ 0.010$&$-0.019$&$-0.032$&$ 0.010$&$ 0.005$\\
$-0.015$&$-0.018$&$-0.040$&$-0.012$&$ 0.002$&$-0.003$&$-0.034$&$-0.013$\\
   OTA07&$-0.010$&$-0.019$&$-0.007$&$ 0.029$&$-0.007$&$-0.012$&OTA00\\
\hline
\end{tabular}
\end{center}
\tablecomments{The table provides \gps\ color terms [mag/mag$(r{-}i)$]
  for each OTA according to its conventional position in GPC1.  These
  offsets do not change for SEDs redder than $(r{-}i)=0.5$.}
\label{tab:otaclr}
\end{table}%

\section{SYSTEMATIC ERRORS}

With more than a hundred high signal-to-noise observations of
spectrophotometric standards and comparisons of thousands of stars
with existing catalogs, the statistical error of this determination of
the \PS\ photometric system is tiny.  In this section we describe our
best estimates of the remaining systematic error, derived both from
the uncertainties in the contributing calculations as well as
whatever external tests we can perform.


The comparison between in-situ measurement of filter transmission and
that performed by Barr only probed one radius, and the match was
excellent but not exact.  We have attempted to compensate for any
spectral tilt and mean, but we estimate that with 90\% confidence the
filter edges are not off by more than 1~nm and the transmission tilt
is not more than $\pm1$\% across any bandpass.  (For example
\rps\ is perhaps slightly more blue sensitive than the Barr curves and
\wps\ slightly more sensitive in the middle of the band.)  Integrating
these limits against power law SEDs yields 3--5 millimag of offset per unit
$(r{-}i)$ from error in band edge and 1--3 millimag from spectral tilt
(\zps\ to \gps).  We therefore estimate the systematic uncertainty in
photometry from imperfect knowledge of filters at 90\% confidence to
be comparable to but smaller than the filter color terms listed in
Table~\ref{tab:filtclr}.

Similarly, the ``tweak'' function corrects the laser-derived
throughput, imposing tilts as large as $\pm3$\% in \gps, and therefore
correcting a color term as large as 10 millimag per unit $(r{-}i)$
relative to Calspec colors.  We do not have any external corroboration
of the accuracy of this correction, but we estimate that it is
accurate enough to bring its contributions to systematic error down to
the same level as that which might be present in the filter curves,
1--3 millimag.

Use of MODTRAN does not alter the fact that we are fundamentally
extrapolating observations of spectrophotometric standards between
airmass 1--2 to other airmass.  In each filter we find an RMS of
$\sim0.01$ mag among $\sim20$ observations of 7 stars distributed more
or less uniformly between airmass 1.0--1.7.  Formally, the uncertainty
in extrapolating to airmass 0 is somewhere around 0.02--0.03 mag,
regardless of whether the extinction was $\sim0.18$ mag per airmass
for \gps\ or $\sim0.04$ mag per airmass for \zps.  The legacy of
that exercise was not a system zeropoint to be applied on different
nights, however, but rather ``top of atmosphere'' $grizyw$ magnitudes for
$3\times10^5$ stars.  These magnitudes are differential measurements
to Calspec spectrophotometric standards taken at the same airmass, and
therefore their formal error is of order 3 millimag, regardless of
filter.  In fact clouds and aerosols can be patchy and do vary on
short timescales, but we believe that the scatter in the standard
observations puts a bound on how large that effect can be.  We
therefore estimate with 90\% confidence that the systematic error
arising from atmospheric extinction is no greater than 5 millimag.

It is well known that the PSF is complex and carries considerable flux
to large angle.  It is typically modeled as a core from atmospheric,
guiding, and optics blurring, followed by a $\theta^{-3}$ skirt from
diffraction, finally succeeded by a $\theta^{-2}$ skirt from small
particle scattering.  This last component generally does not dominate until
larger angle than is used as a ``reference aperture'', but
some 5--10\% of the net flux is scattered beyond any reasonable
aperture, and its loss is normally accounted as a loss in throughput
(dust and degradation) and miniscule enhancement in sky level.
Differential assessment of the fluxes of stars relative to standards
via a single photometry algorithm and reference aperture sidesteps
these PSF issues provided the PSF model does not have biases as a
function of magnitude or PSF shape, except for two purposes.  The
first case arises when comparing stellar photometry to surface
brightnesses of large galaxies, as noted by \cite{TBAD97}.  The second
case arises if we ever try to do absolute photometry and our reference
has different scattering properties than our unknown (perhaps because
it has a different SED or the quantity of dust has changed).  This
change is only visible in PSFs, and a throughput evaluated using a
flatfield or massively defocussed bright star will not detect it.  It
is not inconceivable that the ``tweak'' required to bring standard
star fluxes into agreement with Calspec standards has to do with
chromatic differences in the large angle scattering and systematic
differences in the PSF of blue versus red objects, but it is beyond
the scope of this work to delve deeper into this possibility.

For this exercise we have used a single photometry algorithm, IPP's
PSPhot, restricted to relatively bright objects.  We note that
differences between the flux found by PSPhot, DoPhot, Sextractor, and
other photometry algorithms do exist at the 0.02 mag level, and they
do seem to be related to the ``winginess'' of the PSF.  Also, errors do
enter from the procedure of constructing an aperture
magnitude from a PSF fit magnitude and/or application of a curve of
growth to a fixed metric aperture.  We believe that systematic
errors of at least 10 millimag will arise depending on optics
cleanliness and PSF changes, but most will be taken out by a nightly
regression of flux as a function of airmass.  We believe that the
systematic errors incurred in comparing the PSPhot flux of relatively
bright spectrophotometric standards to others on this particular night
is not larger than 5 millimag.

Our photometric system is based on both direct comparison with the 7
Calspec stars as well as comparing the stellar locus found in the 7
Calspec star fields and MD09 with the stellar locus of all 783 SEDs,
and the agreement provides some level of check on systematic error.
The stellar locus comparison depends on removal of dust reddening,
whose uncertainty we calculated as best we could.  It also depends on
the consistency and homogeneity of the SEDs, but we could not detect
significant differences between the various sources.  By adjustment of
the tweak function we were able to simultaneously match the results
from the 7 Calspec stars to 6 millimag RMS and the cross-filter
stellar locus of three fields, MD09, WD1657, and LDS749b to 10
millimag RMS.  We regard this as confirmation of our 90\% confidence
that our net systematic difference from the 7 Calspec standards is 10
millimag or less.

Although we did not measure absolute fluxes from the laser experiments
nor independently measure the pupil of the telescope, by knowing the
individual throughputs of the optical components and theoretical ray
traces we have created a crude absolute photometer.  If we had
included a contribution for dust or wide angle scattering it is
plausible that we would have decided on a mean loss of 8\% relative to
clean optics.  Although the non-constancy of the tweak function
required to match SEDs was disappointing, it does confirm that these
SEDs are accurately on the AB system within several percent.

The question of how accurately the SEDs conform to the AB system is
complex.  \cite{Bohlin07} describes how the Calspec system is founded
on NLTE models of hot, hydrogen white dwarfs and an absolute flux for
Vega.  We find good consistency among the 7 Calspec stars, although
the fluxes we observe for 1740346 and possibly P177D are lower by
approximately 0.02 mag in \ips\ and \wps\ relative to Calspec than
WD1657 and the three KF stars.  Although our knowledge of \ips\ and
\wps\ may flawed, we also note that there is a discontinuity at 800~nm
where the STIS spectra give way to NICMOS in the Calspec SEDs for
WD1657 and the KF stars, but not for 1740346 and P177D.  Our photometry
may indicate a small discrepancy in some of the Calspec SEDs, but of
course we do not know which is correct.

A more direct comparison of SEDs is also revealing.  \cite{Fukugita96}
list SEDs for Vega and BD+17~4708.  Integrating the SDSS bandpasses
against these and the Calspec SEDs yields $(g{-}z)$ colors that are 24
millimag redder for the SDSS SEDs than the Calspec SEDs.  The NGSL SED
for BD+17~4708 differs very substantially from that of Calspec, with a
difference in $(g{-}z)$ of 87 millimag.  (Although the NGSL data for
BD+17~4708 was subject to a slit mis-center of 0.84 pixels, that is
less than the 0.90 pixel limit for which the web page cautions about
the quality of the V2 correction.)
%

We have no way to know which of these SEDs is in error, although we do
favor the Calspec set because of use of HST, the care with which each
star has been checked, and magnitudes that are usefully faint.  We also
believe that the use of white dwarf models (H and He) will prove to be
superior to subdwarf stars and Vega.  BD+17~4708 is too bright for
\PS\ so we cannot offer support for Calspec versus NGSL, but we do
encourage the community to note and resolve these differences!

Our 90\% confidence estimate for the absolute AB accuracy of the
Calspec set of SEDs is 20 millimag.  We do not believe that it is
presently possible to compare a $g$ magnitude at redshift 0 to a $z$
magnitude at redshift 1 without incurring this level of photometric
uncertainty.

When we compare the \PS\ magnitudes of stars in the MD09 field with
those tabulated by SDSS as part of Stripe82, we find statistically
significant offsets listed in Table~\ref{tab:magtest}.  In particular
$g_{SDSS}-\gps$ is bright by 14 millimag and $r_{SDSS}-\rps$ is faint
by 19 millimag, causing the SDSS $(g{-}r)$ color to be bluer for a
given star than that of \PS\ by $33$ millimag.  We believe that this
may partially arise because of the difference in the SDSS and Calspec
SEDs: if the SDSS standards are redder than Calspec the derived
magnitudes will be bluer.  \cite{SDSS_Doi} has described the evolution
of the SDSS bandpasses over time, and enough change has occurred to
create this level of discrepancy if the SDSS bandpasses we have
adopted from the web page are not correct, since that is how we
transform SDSS magnitudes onto the \PS\ system for comparison.  It is
also possible that the cataloged magnitudes are somewhat heterogeneous
and have acquired offsets from the AB system because of filter
evolution.

\cite{Fukugita11} have performed a detailed comparison of
SDSS catalog magnitudes with synthetic magnitudes and find an offset
$\Delta(g{-}r)_{spec-photo} = 0.026(g{-}r)+0.008$, or $+21$ millimag
in the sense of cataloged magnitudes being bluer than synthetic
magnitudes when evaluated at a common stellar color of $(g{-}r)=0.5$
(close to the discrepancy we see).  We also agree with
\cite{Fukugita11} about the sign and magnitude of the discrepancy in
$(r{-}i)$ (but note the missing minus in their equation for
$\Delta(r{-}i)_{spec-photo}$), and these could both be alleviated by
adjusting $r_{SDSS}$ brighter by about 30 millimag.)
For \cite{Fukugita11} ``This implies that the response curves are well
characterized,'' but we believe that \PS\ and SDSS can do better. 

\begin{figure}[htbp]
\begin{center}
\centerline{\includegraphics[width=\figwid]{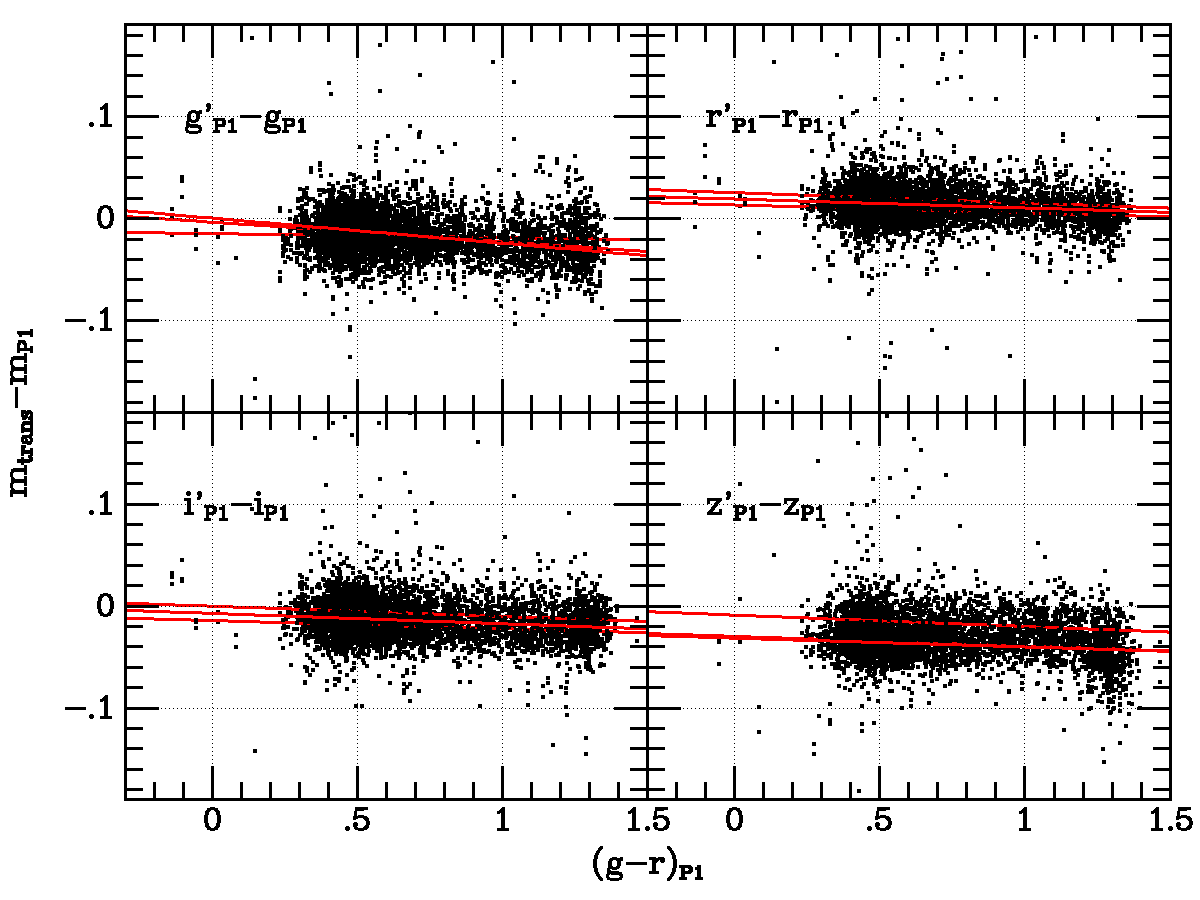}}
\caption{Comparison between stellar magnitudes in MD09 from three
  SDSS releases, transformed to the \PS\ system, and the corresponding
  \PS\ magnitudes.  The Ivezic Stripe82 fits differ particularly for
  \gps\ (small trend with color, but offset from zero) and \zps\ (much
  closer to zero than DR7 and DR8).}
\label{fig:sdssoff}
\end{center}
\end{figure}

As a final comparison we illustrate differences between SDSS DR7
\citep{SDSS_DR7}, SDSS DR8 (Finkbeiner, private communication), and
the Stripe82 compilation from Ivezic in Figure~\ref{fig:sdssoff}.  The
points from the three comparisons are just overlaid, and the lines
illustrate the differences between the three SDSS calibrations.  (We
find that the relations are quite transitive, so these differences
also appear when SDSS is intercompared directly.)
We are therefore inclined to believe that \PS\ is closer to the AB
system than are the extant SDSS catalogs, but the matter deserves
more detailed study.

We do not find Figure~\ref{fig:sdssoff} at all discouraging because
the offsets and slopes are small and very evident, given the quality
of the photometry.  We are confident that the ``ubercal'' procedure
introduced by \cite{Padmanabhan08} and presently being applied to
the 3/4 sky surveyed by \PS\ \citep{Schlafly12} will succeed in
creating an all-sky photometric system with systematic error below 10
millimag.  Merging the SDSS stripes with the \PS\ footprints will help
reduce the errors of both and create a very homogeneous system.

We summarize our best estimates of 90\% systematic uncertainties in
Table~\ref{tab:systematics}.  The most serious systematic
uncertainty comes from the tie between SED and physical units.
\begin{table}[htdp]
\caption{90\% Confidence Systematic Error Estimates [millimag]}
\begin{center}
\begin{tabular}{lcl}
\hline
\hline
Source  &  Uncertainty & Notes\\
\hline
Filter edges             & 3--5  & bigger for broader bandpasses\\
Filter transmission      & 1--3  & bigger for broader bandpasses\\
Tweak determination      & 3--5  & bigger at ends of spectrum\\
Atmospheric extinction   & 3--5  & bigger for bluer wavelengths\\
Flux determination       & 5--10 & inter-night worse\\
Net offset wrt Calspec   & 10    & 7 std, single photometric night\\
SED conformity to AB     & 20    & uncertain\\
\hline
\end{tabular}
\end{center}
\label{tab:systematics}
\end{table}%

\section{SUMMARY}
\label{sec:discussion}

We have described the \PS\ system, comprising telescope, detector, and
software.  Arguing that the photometric properties can be factored
into slowly varying terms (optics, filters, and detector) and rapid
terms (atmosphere), we have endeavored to measure each and to provide
a consistent set of bandpasses and a methodology for determining the
atmospheric transmission.

All optical components and the detector QE were measured separately in
the lab and we measured them in-situ with calibrated, monochromatic
beams of light.  We found good agreement; however, we found that
approximately 8\% of light is lost relative to lab measurements,
presumably because of absorption or scattering by dust and dirt that
have accumulated, and we suspect that one or more lens AR coatings do
not match design.

We have used MODTRAN models to characterize atmospheric transmission.
These are adjusted into agreement with the conditions on a given night
by matching the observed regression against airmass for different
filters to the aerosol content of the model.  We also have deployed a
telescope with full-aperture diffraction grating to monitor the
spectrum of Polaris and constrain the water content of the MODTRAN
models from the equivalent width observed in water bands.

The combination of optics and filter transmission with atmospheric
transmission gives us a net cross-section of the \PS\ system to
convert a photon arrival rate to a detected signal.  For a source
whose AB spectrum is known except for a normalization, we can thereby
invert the observed signal and obtain an absolute AB magnitude.

The comparison with spectral energy distributions was carried out on a
night of exceptional clarity devoted to observations of 7 Calspec
spectrophotometric standards, observed with no filter and in all
filters at a wide range of airmasses.  This comparison revealed the
need for an 0.03 mag RMS ``tweak'' correction to our in-situ
measurements of throughput across the optical whose origin we do not
understand.  By tweaking the in-situ measurements into agreement with
the spectrophotometric standards we obtained transmission functions
for the optics and for each filter, and we have therefore
made the Calspec standards the basis for the \PS\ photometric system.

Given \PS\ bandpasses, we provide a number of useful products, such as
an unreddened stellar locus, Galactic extinction coefficients as a
function of $E(B{-}V)$, and stellar color transformations between
\PS\ and other photometric systems.  We also present the color
terms in the \PS\ system that appear as a function of field angle
among the filters and between the 60 different CCDs.

We finished with a discussion of the (small) random errors and (more
serious) systematic errors that remain in the \PS\ system.  We believe
that we have tied the \PS\ system to the 7 Calspec SEDs to the 10
millimag level or better, but we believe that it is possible that
errors as large as 20 millimag may still exist between the Calspec SEDs
and the AB system.  Comparison with stars cataloged by SDSS reveal
excellent agreement as well as systematic offsets at the $\sim20$
millimag level that we have argued can be traced to systematic errors
in the SDSS bandpasses and systematic differences between SDSS
spectrophotometry and Calspec.

In the future we certainly will obtain observations of more
spectrophotometric standards on photometric nights.  There are
$\sim20$ Calspec stars faint enough not to saturate during
ordinary observing that are particularly useful.

The ``ubercal'' product being generated by \cite{Schlafly12} may also
reveal some interesting systematics while it is creating a homogeneous
catalog of stars around the sky.  In particular we look forward to the
learning how the many epochs of ``ubercal'' magnitudes for the various
Calspec standards match up, as well as the $\sim50$~sq.~deg. observed
on MJD 55744.  It would be worth integrating the SDSS
spectrophotometric SEDs against these \PS\ bandpasses to obtain their
\PS\ magnitudes for comparison with the ``ubercal'' magnitudes.

The ``tweak'' difference between in-situ throughput measurements and
spectrophotometry was disagreeably but not surprisingly large.  It seems
likely that the atmosphere is {\it not} a primary impediment to
squeezing the accuracy of absolute photometry below the 1\% level, and
we could certainly do a much better job with our in-situ measurements,
both relative and absolute.  With some effort it should be possible to
modify our ground-based measurements to the point that they provide
useful constraints on white dwarf models and SEDs measured by HST.  We
look forward to the success of the ACCESS rocket experiments
\citep{ACCESS10} that seek to improve the
absolute calibration of Vega and BD+17~4708.  It is certainly
straightforward to design new, special purpose equipment to do
absolute spectrophotometry from the ground, based on NIST calibration
of photodiodes, that could reach the 1\% level.  Although we have no
immediate plans to carry out such experiments, we emphasize that
knowledge of absolute spectrophotometry is the main limitation in our
current ability to do precision photometry, and we encourage the
community to support efforts to improve it.

{\it Facilities:} \facility{PS1 (GPC1)}

\acknowledgments

Support for this work was provided by National Science Foundation
grant AST-1009749.  The PS1 Surveys have been made possible through
contributions of the Institute for Astronomy, the University of
Hawaii, the Pan-STARRS Project Office, the Max-Planck Society and its
participating institutes, the Max Planck Institute for Astronomy,
Heidelberg and the Max Planck Institute for Extraterrestrial Physics,
Garching, The Johns Hopkins University, Durham University, the
University of Edinburgh, Queen's University Belfast, the
Harvard-Smithsonian Center for Astrophysics, and the Las Cumbres
Observatory Global Telescope Network, Incorporated, the National
Central University of Taiwan, and the National Aeronautics and Space
Administration under Grant No. NNX08AR22G issued through the Planetary
Science Division of the NASA Science Mission Directorate.

\clearpage


\begin{thebibliography}

\bibitem[Abazajian et al.(2009)]{SDSS_DR7} Abazajian, K.~N., 
Adelman-McCarthy, J.~K., Ag{\"u}eros, M.~A., et al.\ 2009, \apjs, 182, 543 

\bibitem[Anderson {\it et al.}(2001)]{MODTRAN} Anderson, G.~P., {\it et al.}, \ 2001 Proc. of the SPIE, {\bf 4381}, 455

\bibitem[Bessell(1990)]{Bessel90} Bessell, M.~S.\ 1999, \pasp, 102, 1181

\bibitem[Bessell(2005)]{Bessel05} Bessell, M.~S.\ 2005,
  \araa, 43, 293

\bibitem[Bessell \& Murphy(2012)]{Bessel12} Bessell, M., \& 
  Murphy, S.\ 2012, \pasp, 124, 140 

\bibitem[Bohlin {\it et al.}(2001)]{Calspec} Bohlin, Dickinson, \&
  Calzetti, 2001, \aj, 122, 2118

\bibitem[Bohlin(2007)]{Bohlin07} Bohlin, R.~C.\ 2007, The Future 
of Photometric, Spectrophotometric and Polarimetric Standardization, 364, 315 

\bibitem[Burke et al.(2010)]{CTIOatmos} Burke, D.~L., Axelrod, T., Blondin, S., et al.\ 2010, \apj, 720, 811 

\bibitem[Chambers {\it et al.}(in prep)]{PS_MDRM} Chambers, K.~C,  {\it et al.}, in preparation.

\bibitem[Cohen et al.(2003)]{Cohen03} Cohen, M., Wheaton, W.~A. \& 
Megeath, S.~T.,\ 2003 \aj, 126, 1090


\bibitem[Covey et al.(2007)]{Covey07} Covey, K.~R., et al.\ 2007, \aj, 134, 2398 

\bibitem[Doi et al.(2010)]{SDSS_Doi} Doi, M. et al.\ 2010, \aj, 139, 1628

\bibitem[Fitzpatrick(1999)]{Fitzpatrick99} Fitzpatrick, E.~L.\ 1999, 
\pasp, 111, 63 

\bibitem[Fukugita et al.(1996)]{Fukugita96} Fukugita, M.,
Ichikawa, T., Gunn, J.~E., et al.\ 1996, \aj, 111, 1748 

\bibitem[Fukugita et al.(2011)]{Fukugita11} Fukugita, M., Yasuda, 
N., Doi, M., Gunn, J.~E., \& York, D.~G.\ 2011, \aj, 141, 47 

\bibitem[Groenewegen(2008)]{Groenewegen08} Groenewegen, 
M.~A.~T.\ 2008, \aap, 488, 935 

\bibitem[Gunn \& Stryker(1983)]{Gunn+Stryker} Gunn, J.E., \&
Stryker, L.L.\ 1983, \apjs, 52, 121

\bibitem[Gunn et al.(1998)]{SDSScam} Gunn, J.~E., Carr, M., 
Rockosi, C., et al.\ 1998, \aj, 116, 3040 

\bibitem[Hayes \& Latham(1975)]{Hayes+Latham75} Hayes, D.~S., \&
  Latham, D.~W.\ 1975, \apj, 197, 593

\bibitem[Hayes(1985)]{Hayes85} Hayes, D.~S.\ 1985, Calibration 
of Fundamental Stellar Quantities, 111, 225 

\bibitem[High {\it et al.}(2009)]{SLR} High, F.~W., Stubbs, 
C.~W., Rest, A., Stalder, B., \& Challis, P.\ 2009, \aj, {\bf 138}, 110. 

\bibitem[Hodapp {\it et al.}(2004)]{PS1_optics} Hodapp, K.~W., Siegmund, 
W.~A., Kaiser, N., Chambers, K.~C., Laux, U., Morgan, J., \& Mannery, E.\ 2004, \procspie, {\bf 5489}, 667 

\bibitem[Kaiser et al.(2010)]{ACCESS10} Kaiser, M.~E., Kruk, 
J.~W., McCandliss, S.~R., et al.\ 2010, \procspie, 7731,  


\bibitem[Kaiser et al.(2010)]{PS1_system} Kaiser, N., et al.\ 
2010, \procspie, 7733,  12K.

\bibitem[Magnier(2006)]{PS1_IPP} Magnier, E.\ 2006, Proceedings of The Advanced 
Maui Optical and Space Surveillance Technologies Conference, Ed.: S. Ryan, The Maui Economic Development Board, p.E5

\bibitem[Magnier {\it et al.}(in prep)]{EMphoto} Magnier, E.,  {\it et al.}, in preparation. 

\bibitem[Oke \& Gunn(1983)]{Oke+Gunn83} Oke, J.~B., \& Gunn, J.~E.\ 1983, \apj, 266, 713 

\bibitem[Onaka {\it et al.}(2008)]{StarGrasp} Onaka, P., Tonry, J.~L., 
Isani, S., Lee, A., Uyeshiro, R., Rae, C., Robertson, L., 
\& Ching, G.\ 2008, \procspie, {\bf 7014},  12.

\bibitem[Padmanabhan et al.(2008)]{Padmanabhan08} Padmanabhan, N., 
Schlegel, D.~J., Finkbeiner, D.~P., et al.\ 2008, \apj, 674, 1217 

\bibitem[Parenago(1940)]{Parenago} Parenago, P.~P.\ 1940, Astron. Zh., 17, 3

\bibitem[Patat {\it et al.}(2011)]{Paranal_atmos} Patat, F., Moehler, S., O'Brien, K., et al.\ 2011, \aap, 527, A91 

\bibitem[Pickles(2011)]{Pickles} Pickles, A.J.\ 1998 \pasp, 110, 863

\bibitem[Schechter et al.(1993)]{DoPhot} Schechter, P.~L., Mateo, M., \& Saha, A.\ 1993, \pasp, 105, 1342 

\bibitem[Schlafly \& Finkbeiner(2011)]{Schlafly11} Schlafly, E.~F., \& 
Finkbeiner, D.~P.\ 2011, \apj, 737, 103 

\bibitem[Schlafly \& Finkbeiner(2012)]{Schlafly12} Schlafly, E.~F., \& 
Finkbeiner, D.~P.\ 2012, in preparation

\bibitem[Schlegel et al.(1998)]{SFD} Schlegel, D.~J., 
Finkbeiner, D.~P., \& Davis, M.\ 1998, \apj, 500, 525 

\bibitem[Shivvers et al.(in prep))]{polar} Shivvers, I. {\it et al.}, in preparation

\bibitem[Simons \& Tokunaga(2002)]{Simons+Tokunaga} Simons, D.~A., \& 
Tokunaga, A.\ 2002, \pasp, 114, 169 

\bibitem[Stubbs \& Tonry(2006)]{PrecisionPhot} Stubbs, C.~W., \& Tonry, J.~L.\ 2006, \apj, 646, 1436 

\bibitem[Stubbs et al.(2007)]{PASPatmos} Stubbs, C.~W., High, 
F.~W., George, M.~R., et al.\ 2007, \pasp, 119, 1163 

\bibitem[Stubbs et al.(2010)]{lasercal} Stubbs, C.~W., Doherty, 
P., Cramer, C., Narayan, G., Brown, Y.~J., Lykke, K.~R., Woodward, J.~T., 
\& Tonry, J.~L.\ 2010, \apjs, 191, 376 

\bibitem[Tonry {\it et al.}(1997)]{TBAD97} Tonry, J. L., Blakeslee, J. P.,
        Ajhar, E. A., \& Dressler, A. 1997, \apj, 475, 399

\bibitem[Tonry {\it et al.}(2008)]{GPC} Tonry, J.~L., Burke,
B.~E., Isani, S., Onaka, P.~M., \& Cooper, M.~J.\ 2008, \procspie,
{\bf 7021}, 9.

\bibitem[York et al.(2000)]{SDSS} York, D.~G., et al.\ 2000, \aj, 120, 1579 

\end{thebibliography}
\end{document}